\documentclass[twocolumn]{aastex63}

\usepackage{gensymb}

\newcommand{\ha}{H$\alpha$}
\newcommand{\hb}{H$\beta$}
\newcommand{\lya}{Ly${\alpha}$}
\newcommand{\sitwo}{SiII}
\newcommand{\ntwo}{NII}

\usepackage{soul,color}
\soulregister\citet7
\soulregister\citep7
\soulregister\ref7
\soulregister\pageref7
\soulregister\lya7
\soulregister\ha7
\soulregister\hb7
\soulregister\sitwo7
\soulregister\ntwo7

\received{May 19, 2021}
\accepted{June 9, 2022}
\submitjournal{ApJ}

\shorttitle{The Lyman Alpha Reference Sample XIII}
\shortauthors{Le Reste et al.}

\graphicspath{{./}}

\begin{document}


\title{The Lyman Alpha Reference Sample XIII:\\ High-Angular Resolution 21cm HI observations of \lya\ emitting galaxies}

\correspondingauthor{Alexandra Le Reste}
\email{alexandra.lereste@astro.su.se, lerestea@gmail.com}

\author[0000-0003-1767-6421]{Alexandra Le Reste}
\affiliation{The Oskar Klein Centre, Department of Astronomy, Stockholm University, AlbaNova, SE-10691 Stockholm, Sweden.}

\author[0000-0001-8587-218X]{Matthew Hayes}
\affiliation{The Oskar Klein Centre, Department of Astronomy, Stockholm University, AlbaNova, SE-10691 Stockholm, Sweden.}

\author[0000-0002-1821-7019]{John M. Cannon}
\affiliation{Department of Physics and Astronomy, Macalester College, 1600 Grand Avenue, Saint Paul, MN 55105, USA.}

\author[0000-0002-8505-4678]{Edmund Christian Herenz}
\affiliation{European Southern Observatory,
  Av. Alonso de C\'ordova 3107,
  763 0355 Vitacura,
  Santiago, Chile}
  
\author[0000-0003-0470-8754]{Jens Melinder}
\affiliation{The Oskar Klein Centre, Department of Astronomy, Stockholm University, AlbaNova, SE-10691 Stockholm, Sweden.} 
\author[0000-0003-4695-6844]{Veronica Menacho}
\affiliation{The Oskar Klein Centre, Department of Astronomy, Stockholm University, AlbaNova, SE-10691 Stockholm, Sweden.} 

\author[0000-0002-3005-1349]{G{\"o}ran {\"O}stlin}
\affiliation{The Oskar Klein Centre, Department of Astronomy, Stockholm University, AlbaNova, SE-10691 Stockholm, Sweden.} 

\author[0000-0003-1111-3951]{Johannes Puschnig}
\affiliation{The Oskar Klein Centre, Department of Astronomy, Stockholm University, AlbaNova, SE-10691 Stockholm, Sweden.}
\affiliation{Argelander-Institut für Astronomie, Auf dem Hügel 71, 53121, Bonn, Germany.} 

\author[0000-0002-9204-3256]{T. Emil Rivera-Thorsen}
\affiliation{The Oskar Klein Centre, Department of Astronomy, Stockholm University, AlbaNova, SE-10691 Stockholm, Sweden.}

\author{Daniel Kunth}
\affiliation{Institut d’Astrophysique de Paris, 98 bis Boulevard Arago, F-75014 Paris, France}

\author{Nick Velikonja}
\affiliation{Department of Physics and Astronomy, Macalester College, 1600 Grand Avenue, Saint Paul, MN 55105, USA.}


\begin{abstract}
The \lya\ emission line is one of the main observables of galaxies at high redshift, but its output depends strongly on the neutral gas distribution and kinematics around the star-forming regions where UV photons are produced. 
We present observations of \lya\ and 21-cm HI emission at comparable scales with the goal to qualitatively investigate how the neutral interstellar medium (ISM) properties impact \lya\ transfer in galaxies.
 We have observed 21-cm HI at the highest angular resolution possible ($\approx3$\arcsec\ beam) with the VLA in two local galaxies from the Lyman Alpha Reference Sample. We contrast this data with HST \lya\ imaging and spectroscopy, and MUSE and PMAS ionized gas observations. 
 In LARS08, high intensity \lya\ emission is co-spatial with high column density HI where dust content is the lowest. The \lya\ line is strongly redshifted, consistent with velocity redistribution which allows \lya\ escape from high column density neutral medium with low dust content. In eLARS01, high intensity \lya\ emission is located in regions of low column density HI, below the HI data sensitivity limit  ($<2\times10^{20}\,$cm$^{-2}$). The perturbed ISM distribution with low column density gas in front of the \lya\ emission region plays an important role in the escape. In both galaxies, the faint \lya\ emission ($\sim1\times10^{-16}$erg.s$^{-1}$cm$^{-2}$arcsec$^{-2}$) traces intermediate \ha\ emission regions where HI is found, regardless of the dust content.
Dust seems to modulate, but not prevent, the formation of a faint \lya\ halo. This study suggests the existence of scaling relations between dust, \ha, HI, and \lya\ emission in galaxies.
\end{abstract}



\section{Introduction} \label{sec:intro}
The \lya\ spectral line is one of the main probes of galaxies in the high-redshift Universe, playing a central role in their detection and redshift determination through either narrow-band filter or spectroscopic observations. The possibility of using this line as a tracer of the first galaxies was initially suggested by \citet{PartridgePeebles1967}: it is indeed one of the strongest intrinsic features in galaxy spectra, with 68\% of ionizing radiation from star-forming regions being reprocessed into the line, assuming case B recombination \citep{Dijkstra2019}. Since the rest wavelength of the line at $\lambda=1215.67$\AA\  is located in the ultraviolet (UV) range, it is redshifted out of atmospheric absorption and becomes observable from ground-based facilities for $z\gtrsim 2$ objects, making the line easier to observe for high-redshift objects. 

Corresponding to the spontaneous transition between the first excited state of the Hydrogen atom and its ground state, the \lya\ line is a resonant transition and is absorbed and then shortly re-emitted by any neutral Hydrogen atom it encounters, resulting in a scattering of the line both spatially and spectrally. Since the rest-frame wavelength of the \lya\ line is located in the UV range it is prone to dust absorption, which effectively removes the absorbed photons from the \lya\ output of the galaxy. These interactions with the interstellar medium (ISM) of galaxies result in a complex radiative transfer of the line, that must be understood for the \lya\ spectra of high-z galaxies to be interpreted correctly and to be used to their maximum potential. However, studying the details of the ISM conditions is not possible in the high-redshift Universe where only global properties can be obtained, due to the loss of resolution and depth suffered by observations. Moreover, no facilities will be built on the current horizon that will detect 21cm emission directly from galaxies at the highest redshifts.

The Lyman Alpha Reference Samples (LARS and eLARS) are two samples containing 42 local (z = 0.028 - 0.18) star forming galaxies that have been assembled to study the parameters impacting the \lya\ radiative transfer in galaxies \citep[][Melinder et al., in prep.]{Ostlin2014,Runnholm2020}. Originally based on Hubble Space Telescope (HST) imaging of the \lya, \ha\ and FUV continuum emission as well as HST Cosmic Origin Spectrograph (COS) spectroscopy, the number of additional observations has grown to include a tremendous amount of supporting data allowing an in-depth investigation of the galaxy properties that influence \lya\ emission. These observations comprise deep optical imaging \citep[][]{Micheva2018}, ionized gas imaging and kinematics using the PMAS \citep[][]{Herenz2016} and MUSE Integral Field Units (IFUs), Infrared observations using SOFIA and Herschel, molecular gas observations with the IRAM 30m telescope and the Atacama Pathfinder EXperiment (APEX) \citep{Puschnig2020} and neutral gas imaging and kinematics with the Green Bank Telescope (GBT) and the Karl G. Jansky Very Large Array (VLA); see \citet{Pardy2014} and Le Reste et al. (in preparation).

The galaxies of the sample have high star formation rates (see Section \ref{sec:sample}) and are close in properties to Lyman Break Galaxies (LBGs), making them analogs to some high redshift galaxies \citep{Hayes2014,Ostlin2014}. Several studies have indeed pointed to compact star forming galaxies as local Universe proxies for high-redshift star-forming galaxies \citep[e.g.,][]{Stanway2014,Henry2015,Izotov2015,Izotov2021}. The LARS and eLARS samples therefore present an unprecedented and comprehensive multi-wavelength view on the key properties driving \lya\ escape that can be reasonably compared to galaxies at higher redshift.

Due to the resonant nature of the line, the \lya\ emission output is expected to depend strongly on a combination of the content, geometry \citep{Giavalisco1996} and kinematics of the HI gas in the galaxy \citep{Kunth1998,Mas-Hesse2003,Wofford2013}. High HI column densities are thought to increase \lya\ scattering and thus the path length of \lya\ photons. In the case where dust is present, this severely increases the probability of \lya\ destruction. 
Therefore processes disrupting the neutral ISM have been proposed to explain the escape of \lya\ photons out of galaxies. Such processes include turbulence (shifting the \lya\ photons out of resonance) or feedback (carving channels in the ISM that lower the column density seen by the affected \lya\ photons); \citep[][]{Keenan2017,Bik2018,Puschnig2020}.
For a review on the parameters impacting \lya\ escape in local galaxies, see \citet[][]{Hayes2019}.

Neutral gas has previously been imaged at low resolution for a subset of the galaxies of the LARS sample ($\sim$ 46\arcsec), thus allowing the comparison of the global HI and \lya\ properties of galaxies \citep[][]{Pardy2014}. However, low-angular resolution observations of 21-cm HI trace the gas on scales that are hardly comparable with those characterising the \lya\ emission at these distances. The neutral medium of the LARS galaxies was also traced with metal absorption lines \citep{Rivera2015}, allowing a comparison of \lya\ emission and the surrounding gas in one aperture for each galaxy. UV absorption lines can be directly compared to \lya\, but most lines trace both the HI and HII gas, and are not unambiguous. For example, CII and SiII are commonly used lines which have ionization potentials of 24.4 eV and 16.3 eV respectively, while HI is ionized at 13.6 eV. Moreover, absorption spectroscopy only provides  information on a single line-of-sight, and thus this technique does not assess the full distribution of neutral medium throughout the galaxy, so that the exact impact of HI distribution on \lya\ emission has never been observed directly. The preferred tracer of the neutral gas medium, the 21cm HI emission line, is very faint compared to e.g \ha\ or \lya\ and so observing it at high angular resolution requires significant surface brightness and long exposure times. Additionally, resolution in the radio domain where the 21cm line is located is inherently lower than in the optical or UV bands, making multi-wavelength comparisons difficult. Therefore even at the comparatively low redshifts of the galaxies in the LARS and eLARS samples, observing the HI 21cm emission remains a challenge.

This study presents the first direct comparison of the resolved HI 21-cm and \lya\ emission lines in galaxies. We have obtained VLA data of the HI line at the highest angular resolution possible ($\sim$3\arcsec\ beam) for two galaxies, enabling the study of the two emission lines on scales that are comparable. We also use a wealth of additional data including IFU ionized gas observations and molecular gas data that allows the tracing of star formation regions where \lya\ is produced and dust absorbing \lya. By examining the interstellar medium phases responsible for the creation, destruction and scattering of \lya\ photons, this study depicts a holistic view of the \lya\ radiative transfer in star-forming galaxies, and tests current theories for \lya\ escape mechanisms.

We present the sample used for this study in section \ref{sec:sample}. The observations and data reduction procedures are described in section \ref{sec:observations}. In section \ref{sec:results} we present the HI data products and the main results obtained through the comparison of various ISM phases and \lya\ emission. In section \ref{sec:discussion} the results are discussed and we provide a summary and conclusion in section \ref{sec:conclusions}.\\

\section{Sample selection} \label{sec:sample}

\begin{deluxetable*}{ccccccccccc}[t]
\tabletypesize{\scriptsize}
\tablenum{1}
\tablecaption{Properties of the galaxies in the high angular resolution HI sample.\label{tab:sample}}
\tablewidth{0pt}
\tablehead{
\colhead{ID} & \colhead{RA} & \colhead{DEC} & \colhead{z} & \colhead{D} & \colhead{M$_*^{a}$} & \colhead{SFR $^{a}$} & \colhead{12 + log(O/H)$^{a}$} & \colhead{L$_{\textrm{\lya}}\,^{a}$} & \colhead{EW$_{\textrm{\lya}}\,^{a}$} & \colhead{f$_{esc}(\textrm{\lya})^{a}$}\\
& & & & [Mpc] & [10$^{10}$ M$_{\odot}$] & [M$_{\odot}$/yr] & & [10$^{41}$erg s$^{-1}$]& [\AA]& [\%]\\
(1) & (2) & (3) & (4) & (5) & (6) & (7) & (8) & (9) & (10) & (11) 
}

\startdata
LARS08 & 12h50m13.5s & +07d34m42s & 0.038 & 168.3 & 11.4$\pm$ 0.12 & 60.53  $\pm$ 0.28 & 8.565$\pm^{0.011}_{0.013}$ & 3.94 $\pm$ 0.24 & 17.02 $\pm$ 0.53 & 0.40 $\pm$ 0.02 \\
eLARS01 & 16h11m40.8s & +52d27m24s& 0.029 & 129.0 & 7.98$\pm$0.05  & 24.70  $\pm$ 0.03 & 8.758$\pm^{0.009}_{0.015} $ & 5.50 $\pm$ 0.08 & 22.70 $\pm$ 0.35 & 1.37 $\pm$ 0.03 \\
\enddata
\tablecomments{Notes on columns: (1) - Galaxy ID. (2) - Right Ascension (J2000). (3) - Declination (J2000). (4) - Redshift obtained from nebular line fitting of the SDSS spectra. (5) - Luminosity distance calculated from the redshift assuming $\Lambda$CDM cosmology with parameters $\Omega_m=0.3$, $\Omega_d=0.7$, $H_0=70$ km s$^{-1}$ Mpc$^{-1}$ . (6) - Stellar Mass. (7) - Star Formation Rate. (8) - Metallicity. (9) - \lya\ Luminosity. (10) - \lya\ equivalent width. (11) - \lya\ escape fraction.\\
$^{a}$ Melinder et al. in preparation.}
\end{deluxetable*}

The two galaxies in this study were drawn from the LARS and eLARS galaxy samples, aiming to investigate the physical properties that impact \lya\ emission in local galaxies. The LARS sample contains 14 nearby starburst galaxies ($0.028\leq z \leq 0.18$) that were selected for their strong star formation by their \ha\ equivalent width (EW$_{\textrm{H}\alpha} \geq 100 $ \AA) and FUV luminosity ($9.5\leq\log (\textrm{L}_{FUV}/\textrm{L}_{\odot})\leq 10.7 $) \citep[see][for more details]{Ostlin2014,Hayes2014}. It was later extended with the eLARS sample, containing 28 additional galaxies that evenly sample the \ha\ equivalent width and FUV luminosity parameter spaces (Melinder et al., in prep.). Preliminary fluxes for the eLARS sample have been presented in \citet{Runnholm2020}. With lower \ha\ equivalent width constraints (EW$_{\textrm{H}\alpha}\geq 40$ \AA), the eLARS galaxies are  less extreme in their properties and are representative of the galaxy population dominating the local FUV luminosity function. Together the LARS and eLARS samples are suited to comprehensively study the parameters that impact the radiative transfer of Ly$\alpha$ photons in local galaxies.\\

We have obtained VLA D and C configuration observations for nearly all the galaxies in the samples and B configuration for 15 of them; these observations will be presented in future work (Le Reste et al., in prep.). While the LARS and eLARS samples were selected to contain galaxies
comparable to those identified in UV-based surveys at high-redshift \citep[e.g.][]{Cowie2011}, the
targets selected for the highest angular resolution 21cm emission followup (VLA A configuration) make up a small subset of only two galaxies: LARS08 and eLARS01. Basic properties of the galaxies in the subsample are presented in Table \ref{tab:sample}. In order to obtain the highest quality HI data with long baselines, we began our campaign targeting the galaxies with the largest 21cm flux integrals based upon VLA D and C imaging. This selection also leads to targeting more massive and gas-rich galaxies, which may also be more metal and dust-rich and thus have lower \lya\ output.  This can be seen from the data presented in Table \ref{tab:sample}, which shows that while the luminosities and EWs are sufficient for them to be identified in high-$z$ \lya\ surveys, the global \lya\ escape fractions are rather low at $\sim 1$~\%. Nevertheless, we stress that the emission physics, which depends significantly upon HI and dust will be identical in all galaxies.

\section{Observations and data reduction} \label{sec:observations}

\subsection{VLA Observations}
The galaxies in the sample were observed with the D, C, B and A configurations of the Very Large Array\footnote{The VLA is operated by the National Radio Astronomy Observatory.} (project numbers 13A-181, 14A-077, 17A-240 and 18A-095). Observations made use of the L-band centered on the galaxies respective redshifted 21-cm HI frequency. Information about the observations for each array configuration can be found in Table \ref{tab:obs}. Lower angular resolution observations of LARS08 were previously presented in \citet{Pardy2014}. We designed our observations to reach the highest possible angular resolution by following the guidelines spelled out by the NRAO\footnote{https://science.nrao.edu/facilities/vla/docs/manuals/oss/performance/comb-conf-mosaicking}. These guidelines specify integration times should increase by a factor of $\sim$3 when stepping through the longer array configurations to reach homogeneous sensitivity limits; since we are concerned with HI emission, such common limits would correspond to a fixed gas column density. Since designing the observations we have become aware of the fact that these guidelines may be most appropriate for relatively compact emission, while in our case some of the HI gas – especially the warm neutral phase, tidal debris, and circumgalatic gas – may be extended. Nevertheless, we believe this strategy is still appropriate for several reasons. Firstly, we do not attempt to use the A and B configuration data to study the large-scale emission for which we rely solely upon imaging of the D+C data which has higher sensitivity. The observed galaxies are at distances of 129 and 168 Mpc, which are large compared to the distances of most galaxies with interferometric HI observations. Thus our physical resolution elements are comparatively coarse (2-3 kpc at half power) in comparison to the expected size of HI clouds. Thus we expect these structures to appear compact and point-like, which would make the choice of integration times appropriate. We stress that the A-configuration data is used only when we make the comparisons on the smallest scale (i.e. with UV/optical data) and that in these cases we are likely sensitive only to the highest column density, clumpiest gas.

\begin{deluxetable}{ccccc}[h]
\tablenum{2}
\tablecaption{Observation parameters for the high angular resolution LARS VLA sample.\label{tab:obs}}
\tablewidth{0pt}
\tablehead{
\colhead{ID} & \colhead{Conf.} & \colhead{Date} & \colhead{$T_{int} [h]$} & \colhead{Calibrator}  
}

\startdata
L08  & D& Apr 2013 & 2.18 & J1254+1141  \\
  &  C&  Jun 2017 & 5.87&  $\vert$ \\
  &  B& Nov 2017 - Jan 2018 & 12.95 & $\vert$ \\
  &  A& May - Jun 2018 & 37.44 & $\vert$ \\
eL01 &  D & Aug 2014 & 1.63 & J1035+5038  \\
  &  C& Jun 2017 & 5.81 & $\vert$  \\
  &  B& Sep - Dec 2017 & 12.38 & $\vert$  \\
  &  A& Mar - Jun 2018 & 36.40 & $\vert$ \\
\enddata
\tablecomments{Notes on columns, from left to right: (1) - Galaxy name. (2) - VLA array configuration. (3) - Observations date. (4) - Integration time. (5) - Source used as phase calibrator.}
\end{deluxetable}

\subsection{HI Data reduction}
\subsubsection{RFI masking and Calibration}
The VLA HI data was reduced with the \texttt{casa} software using standard prescriptions. Radio-Frequency Interference (RFI) were removed from the data to avoid the introduction of imaging artifacts in the final data cubes. A two step procedure was applied to identify and exclude RFI. First, the automatic RFI flagging task \texttt{tfcrop} was applied with parameters \texttt{maxnpieces}=3 \texttt{timecutoff}=3.0 and \texttt{freqcutoff}=3.0. This led to the removal of $\sim$1\% of the interferences in the time-frequency plane, consisting mainly of strong, frequency-narrow RFI. This was followed by eye inspection (using the \texttt{plotms} and \texttt{viewer} tasks) and hand-masking of the measurement sets using the \texttt{flagdata} task. The resulting RFI-free measurement sets were then calibrated using 1331+305=3C286 as the bandpass calibrator and either J1254+1141 or J1035+5038 as the phase calibrator. The measurement sets containing all sources were split using the task \texttt{split} to keep only the visibilities corresponding to the target galaxies. The calibrated target-only measurement sets were continuum subtracted by fitting a polynomial of order 1 on line-free channels of each individual dataset using the task \texttt{uvcontsub}. The visibilities of each dataset were reweighted according to their scatter using the task \texttt{statwt} on the same channels as those where \texttt{uvcontsub} was applied, a necessary step when combining data from different array configurations. Typically, we used 200 line-free channels on each side of the HI line, making sure not to include channels at the ends of the bandpass. 

\subsubsection{Cleaning}
The resulting datasets were cleaned to 0.5$\sigma$ using the CASA task \texttt{tclean} in interactive mode with the \texttt{auto-multithresh} algorithm \citep[][]{automultithresh2020} to identify regions containing emission. This was complemented by manual inspection of the clean region selected. The use of \texttt{auto-multithresh} allowed for the detection of a small HI object at the edge of the LARS08 field that was missed by eye identification due to the size of the object and remote location from the center of the field. It also helped speeding the clean region selection process considerably, especially in the high-angular resolution cubes. The clean images were set to have a common beam in the \texttt{tclean} task and a Briggs weighting robust parameter of 0.5 was used for each individual configuration, combining low and high angular resolution configurations together (D+C, D+C+B and D+C+B+A). The spectral channel width of the cubes was set to 10 km\,s$^{-1}$. Additionally, a uv-taper of 50k$\lambda$ was used for the D+C+B+A cubes to increase the flux sensitivity while keeping a reasonably high angular resolution. Table \ref{tab:hi} summarizes the properties of each data cube.

\subsubsection{Emission detection}
After data reduction, the emission regions were isolated from the noise. For that purpose, the cubes were masked with dilated masking, using an algorithm developed for the identification of CO emission in local galaxies \citep{Rosolowsky2006,Sun2018}. The algorithm first identifies regions above a certain signal-to-noise ratio using an intensity threshold criterion \texttt{snr$_{hi}$} and a criterion on the number of consecutive channels \texttt{n$_{hi}$} for which the intensity threshold must be seen. This first mask is then extended spatially and spectrally to include regions with a lower intensity threshold \texttt{snr$_{lo}$} found across a certain number of consecutive channels \texttt{n$_{lo}$}. This type of emission detection method allows for fainter emission regions to be detected, assuming that they are connected to the higher intensity emission regions, and also limits the noise included in the final data products. After conducting tests on several datasets, we found that the parameter combination (\texttt{snr$_{hi}$}=3.5-3, \texttt{n$_{hi}$}=3, \texttt{snr$_{lo}$}=1.5-2, \texttt{n$_{lo}$}=1) and extending the mask spatially by 2 pixels allowed for the best emission recovery in our datasets while limiting noise inclusion. The masked cubes were primary beam corrected to obtain correct flux densities for the subsequent data products derivation.

\subsection{HI data products}
\subsubsection{HI flux, mass, spectra}
To obtain the flux and mass associated with the HI emission, the masked, primary beam corrected cubes for the VLA D array configuration only were used. This choice is motivated by the D configuration having the highest surface brightness sensitivity, thus enabling the detection of HI flux on extended scales. Furthermore the nominal D configuration beam size at 21cm ($\sim40$") is well-matched to the angular sizes of the galaxies as shown in Figure \ref{fig:mom_full}. Although it also contains information from the D array, the D+C+B+A configuration data is more sensitive to higher surface brightness HI due to the lower sensitivity of our longer baseline observations. Thus large-scale emission is resolved out, leading to a loss of flux.\\

The flux S$_{HI}$ in a spectral cube containing a certain number N of pixels $i$ with flux density $I_i$ can be expressed as:
\begin{equation}\label{eq:hi_flux}
\left(\frac{\textrm{S}_{HI}}{\textrm{Jy.km s$^{-1}$}}\right) =  \frac{4 \log(2) {\ell_{pix}}^2}{\pi \theta_{min} \theta_{maj}}\sum_{i}^{N} \left(\frac{I_i}{\textrm{Jy/beam}}\right) \left(\frac{\Delta\textrm{v}}{\textrm{km s$^{-1}$}}\right)
\end{equation}
with $\theta_{min}$ and  $\theta_{maj}$ the beam minor and major axis in degrees, $\ell_{pix}$ the angular length of a pixel in degrees and $\Delta\textrm{v}$ the channel width. The error on the flux was derived accounting for the 3\% typical flux error on the primary calibrator \citep{Perley2017} and the noise in the cube.
Assuming that the gas is optically thin, the HI gas mass can be derived from the flux using the following expression, where D is the luminosity distance:
\begin{equation}\label{eq:hi_mass}
\left(\frac{\textrm{M}_{HI}}{\textrm{M}_{\odot}}\right) = 2.36\times10^5  \left(\frac{\textrm{D}}{\textrm{Mpc}}\right)^2 \left(\frac{\textrm{S}_{HI}}{\textrm{Jy.km s$^{-1}$}}\right)
\end{equation}
Spectra were obtained by summing the flux in each spectral channel of the masked, primary beam corrected D configuration data. We estimate the typical r.m.s. in adjacent frequency channels devoid of 21cm flux by collapsing the mask along the spectral direction, and computing the total noise level.

\subsubsection{Moment maps}
Moment maps were derived using the masked, primary beam corrected D+C+B+A cubes with the \texttt{spectral\_cube} Python package. Both cubes have an average synthesized beam size of 3\farcs6.
Moment-0 maps (integrated emission maps) are derived using:
\begin{equation}\label{eq:mom0}
\textrm{M}_{0} = \sum_{v} I_{v} \Delta\textrm{v}
\end{equation}
Moment-1 maps are intensity-weighted velocity maps, corresponding to the velocity centroid across the spectral axis:
\begin{equation}\label{eq:mom1}
\textrm{M}_{1} = \frac{\sum_{v} v\, I_{v} \Delta\textrm{v}}{\textrm{M}_0}
\end{equation}
Moment-2 maps are linked to the velocity dispersion across the spectral axis, and can be expressed as:
\begin{equation}\label{eq:mom2}
\textrm{M}_{2}=\frac{\sum_{v} (v-\textrm{M}_1)^2\, I_{v} \Delta\textrm{v}}{\textrm{M}_0}
\end{equation}
Velocity dispersion maps can be derived from the moment-2 maps using $\sigma = \sqrt{\textrm{M}_{2}}$, although this method can sometimes lead to overestimating the true velocity dispersion \citep[][]{Mogotsi2016}.

\begin{deluxetable*}{clccccccc}[t]
\tablenum{3}
\tablecaption{HI cube properties\label{tab:hi}}
\tablewidth{0pt}
\tablehead{
\colhead{ID} &\colhead{Conf.} &
\colhead{Beam} & \colhead{rms} & min N$_{HI}$ & \colhead{scale} & S$_{HI}$ &  M$_{HI} $ & z$_{HI} $
 \\
 & & (bmin,bmaj,pa) & [mJy/beam]&[cm$^{-2}$]& [kpc/$\overline{beam}$] & [Jy km s$^{-1}$] & [10$^{10}$ M$_{\odot}$]
 }

\startdata
LARS08 &  D& 45.2",56.6",4.5° & 0.74 & 6.30$\times$10$^{18}$  & 41.5 & 1.75$\pm$0.09 &1.17$\pm$0.06  & 0.03822 \\
  &   D+C& 16.9",18.6",31.6° & 0.40 & 2.77$\times$10$^{19}$& 14.5 & 2.34$\pm$0.08 & 1.56$\pm$0.06 & \\
    & D+C+B   & 6.45",7.09",53.6° & 0.24& 8.57$\times$10$^{19}$ & 5.5 & 2.07$\pm$0.08 & 1.38$\pm$0.05 & \\
    & D+C+B+A  & 3.29",3.95",-85.3° & 0.16 & 2.0$\times$10$^{20}$& 2.9 & 1.35$\pm$0.06 & 0.9$\pm$0.04  & \\
eLARS01 &  D &  42.2",53.4",67.5° & 0.85 & 8.22$\times$10$^{18}$& 29.9 & 4.3$\pm$0.21 & 1.69$\pm$0.08   & 0.02948\\
  &  D+C& 15.6",16.2",-43.1° & 0.42 & 3.62$\times$10$^{19}$ & 9.9 & 2.93$\pm$0.1 &  1.15$\pm$0.04 &  \\
    &  D+C+B & 5.3",7.1",-89.0° & 0.28 & 1.21$\times$10$^{20}$& 3.8 & 2.78$\pm$0.1  & 1.09$\pm$0.04 & \\
    &  D+C+B+A & 3.1",4.1",-80.7° & 0.17 & 2.18$\times$10$^{20}$& 2.2 & 1.69$\pm$ 0.07 &  0.67$\pm$0.03 & \\
\enddata
\tablecomments{Notes on columns, from left to right: (1) - Galaxy ID. (2) - Array configuration data used to create the cube. (3) - Synthesized beam. (4) - Cube rms. (5) - Limiting column density. These correspond either to the 2$\sigma$ or 1.5$\sigma$ limit, depending on the \texttt{snr$_{lo}$} level specified during masking. (6) - average physical diameter of the beam. (7) - HI flux.  (8) - HI mass.  (9) - Redshift obtained from 21cm line centroid.}
\end{deluxetable*}

\subsection{Additional data}
\label{sec:other_data}
\subsubsection{HST data}
HST \lya\ images were obtained using the Solar Blind Channel (SBC) of the Advanced Camera for Surveys (ACS), using the F125LP and F140LP filters to isolate \lya\ and the adjacent continuum, respectively. For details of the observations and data-processing, see \citet{Hayes2014} and \citet{Ostlin2014}; the full procedure will soon be described in \citet[][in prep]{Melinder2022}. In short, we use two adjacent long-pass filters in the SBC to isolate \lya\ and subtract the stellar continuum, using a spatially resolved spectral modeling method relying upon 3 additional optical HST images and spectral synthesis models.

We also use HST spectroscopic data from the Cosmic Origins Spectrograph (COS): the LARS\,08 spectrum was obtained as part of our own program and is presented in \citet{Rivera2015}, while the eLARS\,01 spectrum was presented in \citet{leitherer2013}.  In both cases \lya\ was observed through the G130M grating, which provides a nominal spectral resolution of $R \approx 17000$ at the  wavelength of \lya.  The eLARS\,01 data were re-processed with the COS pipeline following the methods described in  \citet{Rivera2015}.\\

\subsubsection{MUSE data}
Optical integral field spectroscopy of LARS08 was obtained from the Multi-Unit Spectroscopic Explorer \citep[MUSE;][]{Bacon2010} at the Very Large Telescope Unit Telescope 4 under ESO programme 0101.B-0703.  Observations were obtained on 18 May 2018, under a DIMM seeing of $\approx 0.9$ arcsec at full-width-half-maximum. We obtained four individual exposures, each of 650 seconds and rotated by 90 degrees between integrations.  An independent sky frame (120 seconds) was also taken in order to improve the background subtraction. Data were reduced by the standard MUSE pipeline Version 2.8.3 \citep[][]{MUSEpipeline}.\\

 The reduced cube was continuum subtracted using a 151 pixel running median filter in each spaxel. We then fit the fluxes and velocity parameters for the nebular gas using the approximation of single-component Gaussian functions: for every spatial pixel, we simultaneously fit a large number of optical lines ($\simeq$ 40).  We adopt a method that recovers a single common recession velocity and velocity dispersion for all spectral features (accounting for the variation of the MUSE spectral resolution as a function of observed wavelength), and only the normalization of each line  (the flux) is allowed to vary for the lines individually.  In practice this allows all lines to contribute to the velocity measurements at low SNR; at high SNR the strongest lines of hydrogen and
[O III] dominate in the recovery of the kinematic properties, which in turn constrains the profile shape of the weaker lines. This effectively removes degeneracies in the fitting of weaker lines, and allows their flux to be better recovered.\\

\subsubsection{PMAS data}
PMAS Lens-Array \citep{Roth2005} observations for eLARS01 were
obtained under overcast conditions on March 31st, 2016 (DIMM seeing $\sim$2.0"). The
observation is part of an integral field spectroscopic campaign to
study the ionised gas kinematics of the combined LARS+eLARS sample
(Herenz et al., in prep.).  We used the 16$\arcsec\times16\arcsec$ field sampled
at 1$\arcsec\times1\arcsec$ for each spectral pixel.  Moreover, we used the
backwards-blazed R1200 grating and read out the CCD unbinned along the
dispersion axis.  The exposure time was 1800s. No separate sky
exposure was taken, as the targeted \ha\ emission line was not
impacted by telluric emission.  We reduced the data with the p3d
pipeline \citep{Sandin2010} following the same steps as in
\cite{Herenz2016}, except that we did not sky-subtract and flux
calibrate the data, as neither of those steps is required
for the extraction of kinematic information.  Measuring the FWHM of
the calibration arc lamp lines we obtain $R \approx 5300$ as the final
resolving power (or $v_\mathrm{FWHM} = 57$ km\,s$^{-1}$) of the PMAS
data at the observed wavelength of H$\alpha$. Other \ha\ kinematics observations of eLARS01 can be found in \citet[][]{Sardaneta2020}.

\section{Results} \label{sec:results}

\subsection{HI Properties}

\begin{figure*}[!t]
    \centering
    \includegraphics[width=0.97\textwidth]{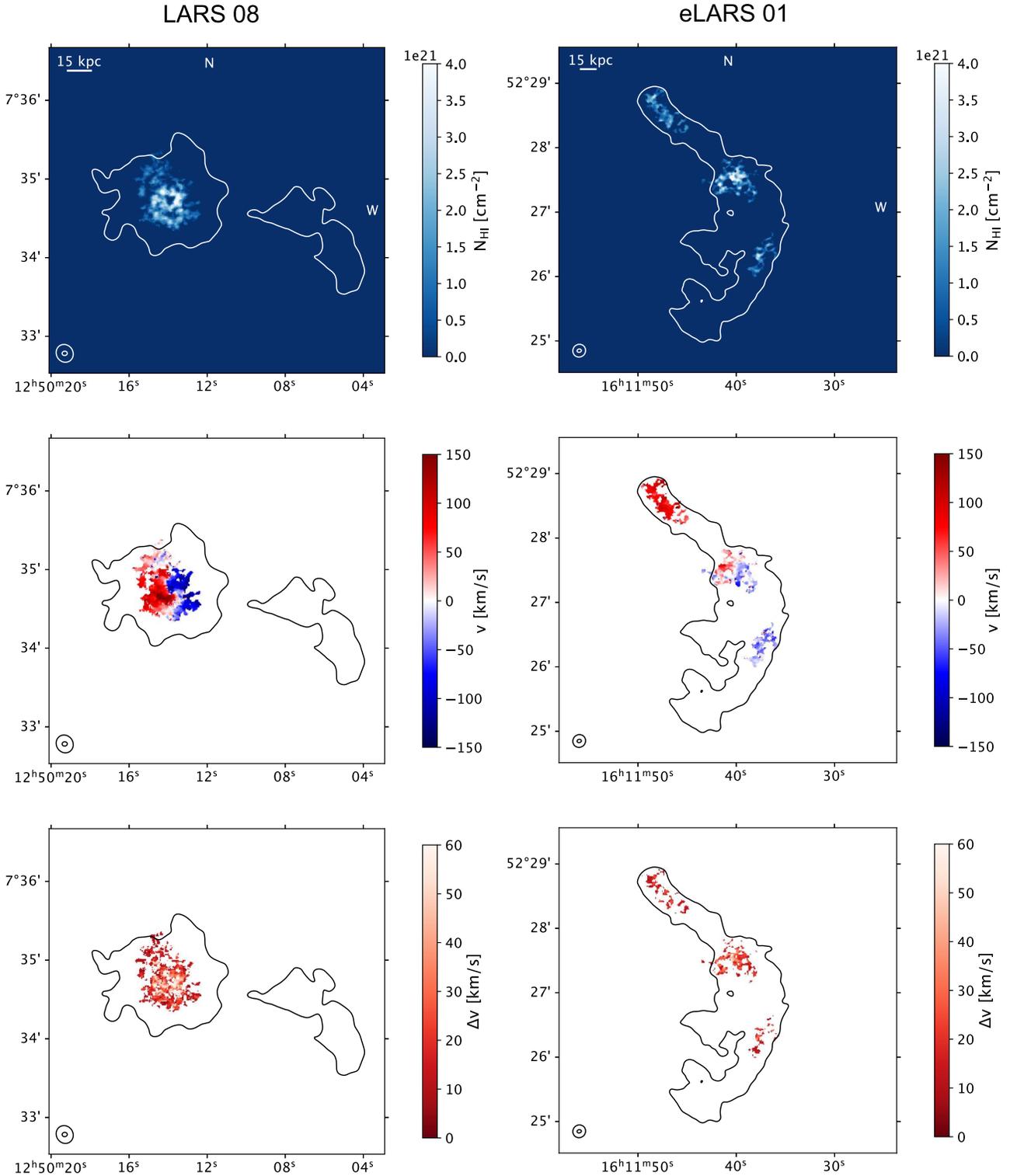}
    \caption{Moment maps for the 21cm HI emission line of LARS08 (left) and eLARS01 (right), obtained with the VLA D+C+B+A configurations. White or black contours characterizing the extent of the D+C data are overlaid ($N_{HI}=5\times10^{19}$cm$^{-2}$). From Top to bottom: integrated intensity (moment-0) map in column density, rest-frame intensity-weighted velocity (moment-1) map, velocity dispersion map. The size of the synthesized beams for the D+C and D+C+B+A data are indicated on the bottom left of the maps by a white or black ellipse. North and West direction are indicated on the moment-0 maps.}
    \label{fig:mom_full}
\end{figure*}
Moment maps of the HI emission in the D+C+B+A cubes for LARS08 (left) and eLARS01 (right) are presented in Figure \ref{fig:mom_full}, along with black or white contours showing the extent of the emission in the D+C cubes. Moment maps for the D, D+C, D+C+B and D+C+B+A are shown in Appendix Figures \ref{fig:moment_lowres_L08} and  \ref{fig:moment_lowres_eL01}. We present channel maps for the D+C+B+A data in Figures~\ref{fig:l08_hi_cm_full} and \ref{fig:el01_hi_cm_full}. The extent of the 21cm emission differs when probed by the different array configurations. This is due to the most compact configuration data (D,C; low-resolution data) having better surface brightness sensitivity, and thus being better suited at detecting faint emission. 
Additionally, large-scale emission is resolved out by the most extended configuration arrays (B,A) due to the lack of short spacing between antennas. The combination of different array configuration data thus yields a compromise between brightness sensitivity and angular resolution that is sufficient to image the high surface brightness 21cm emission at high angular resolution in the regions where \lya\ is produced.\\

The HI emission of both galaxies shows signs of galaxy interaction. For LARS08, the D+C contours indicate the presence of an  HI component South West of the center of the galaxy, with an extended HI feature resembling a tidal tail likely due to a recent infall. The moment maps in Figure \ref{fig:moment_lowres_L08} also show they are connected in velocity-space. The D+C+B+A data, which probes gas with higher column density, does not show the companion object. Inspecting optical data from SDSS \citep{SDSSdr16} and the Pan-STARRS1 survey \citep{Panstarrs}, two extremely faint optical objects were found on the Western edge of the companion HI object, at projected angular distances of 14 arcsec and 7 arcsec from the center of the HI cloud. Due to their offset position and their faintness, we could not unambiguously identify either of these two objects as the counterpart of the HI companion. Another small, faint HI object was found in the VLA field at a distance of 13.2 arcminutes to the center of LARS08, with no detected optical counterpart. Due to the large distance of this object to the galaxy and the lack of direct evidence for its connection, we did not extend the area shown in the figure to display it, but note this suggests that LARS08 is in a group environment.

The HI emission of LARS08 traces a prominent spiral arm in the galaxy and resembles a rotating disk extending beyond the optical diameter of the galaxy. An HI hole is present in the center of the galaxy. This feature linked to the presence of star forming regions is commonly observed in galaxies, with atomic hydrogen transitioning to a molecular phase in the center of the HI disk \citep[see e.g.][]{Wong2002,Blitz2004,Murugeshan2019}. The highest HI column density region of the galaxy ($N_{HI}> 4\times10^{21}\,$cm$^{-2} $) is found immediately to the West of the HI hole.\\

eLARS01 shows a merger-like HI morphology, with tails of HI extending to the North and South of the galaxy that extend on scales of $\sim 100$ kpc each. The emission in the tails can be better appreciated in the D+C contours and moment maps of the lower angular resolution data presented in Figure \ref{fig:moment_lowres_eL01}. Most of the emission in the tails is not recovered with the D+C+B+A data despite the 36 hours of integration with the VLA A configuration. This is probably due to the large angular scale over which the HI emission is distributed and hence a large fraction of the HI gas is found in an extended component that is resolved out in the D+C+B+A configuration data. The decrease in flux recovered between the D configuration-only and the D+C+B+A configuration data also supports this hypothesis (see Table \ref{tab:hi}).
The central part of the galaxy shows, like in the case of LARS08, a central HI hole neighboring the highest intensity region.\textbf{ However, careful examination reveals the presence of an absorption feature centered on the core of one of the merging galaxies (see Appendix Figure \ref{fig:el01_abs}}. \\

Comparable HI morpho-kinematics to those of LARS08 and eLARS01 have been seen in low angular-resolution observations of some of the LARS sample galaxies \citep[see e.g. LARS03 in][]{Pardy2014}, other nearby \lya-emitting starburst galaxies (i.e. the extended HI tails of Tol 1924-416 in \citet[][]{Cannon2004}, see also \citet{Hayes2005, Ostlin2009}), and are common in local dwarf star-forming galaxies \citep{Jaiswal2020}. This is consistent with galaxy mergers playing a role in triggering starburst episodes in galaxies.\\
{\begin{figure}[!b]
    \centering
    \includegraphics[width=0.49\textwidth]{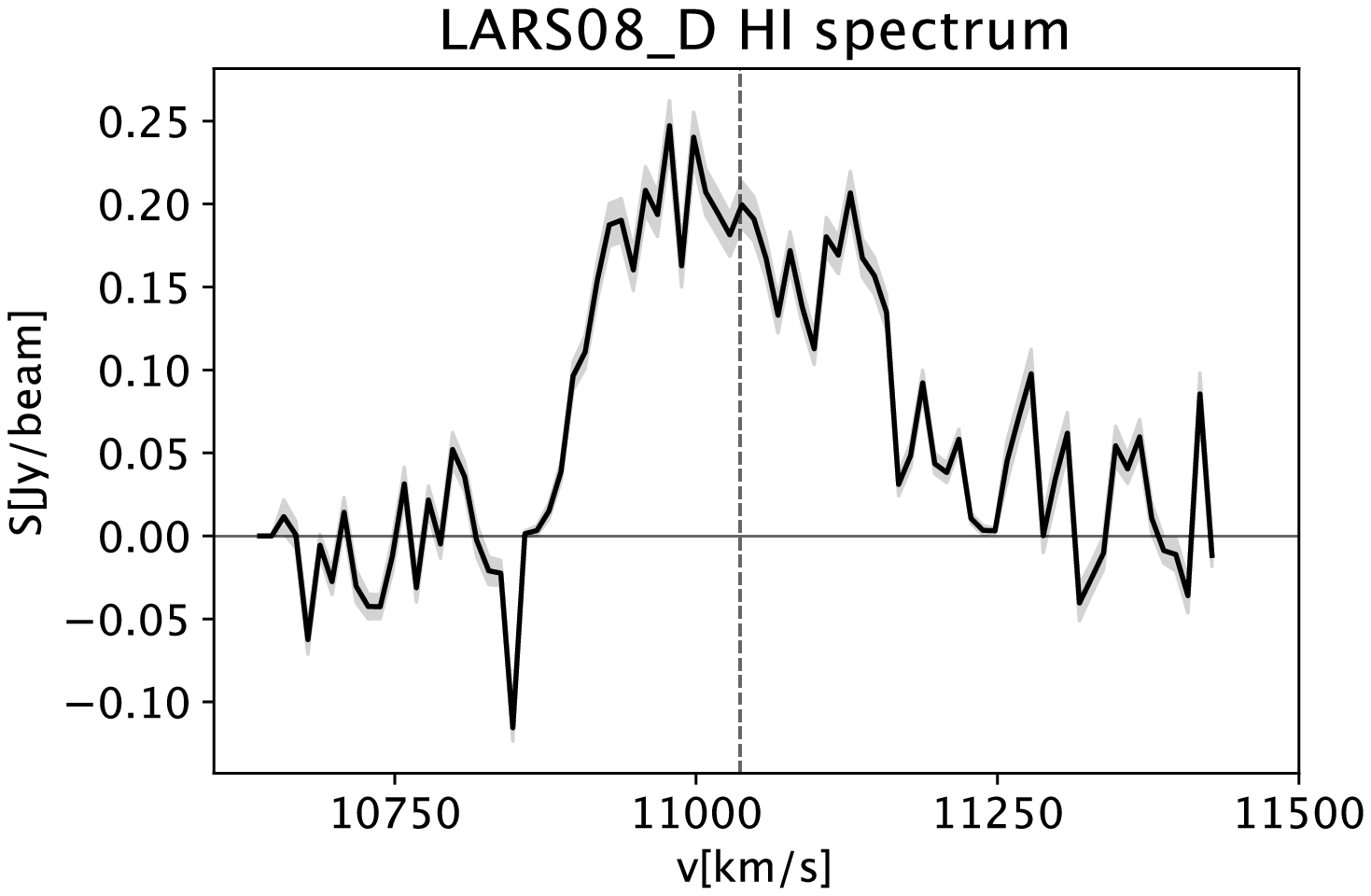}\\
    \includegraphics[width=0.49\textwidth]{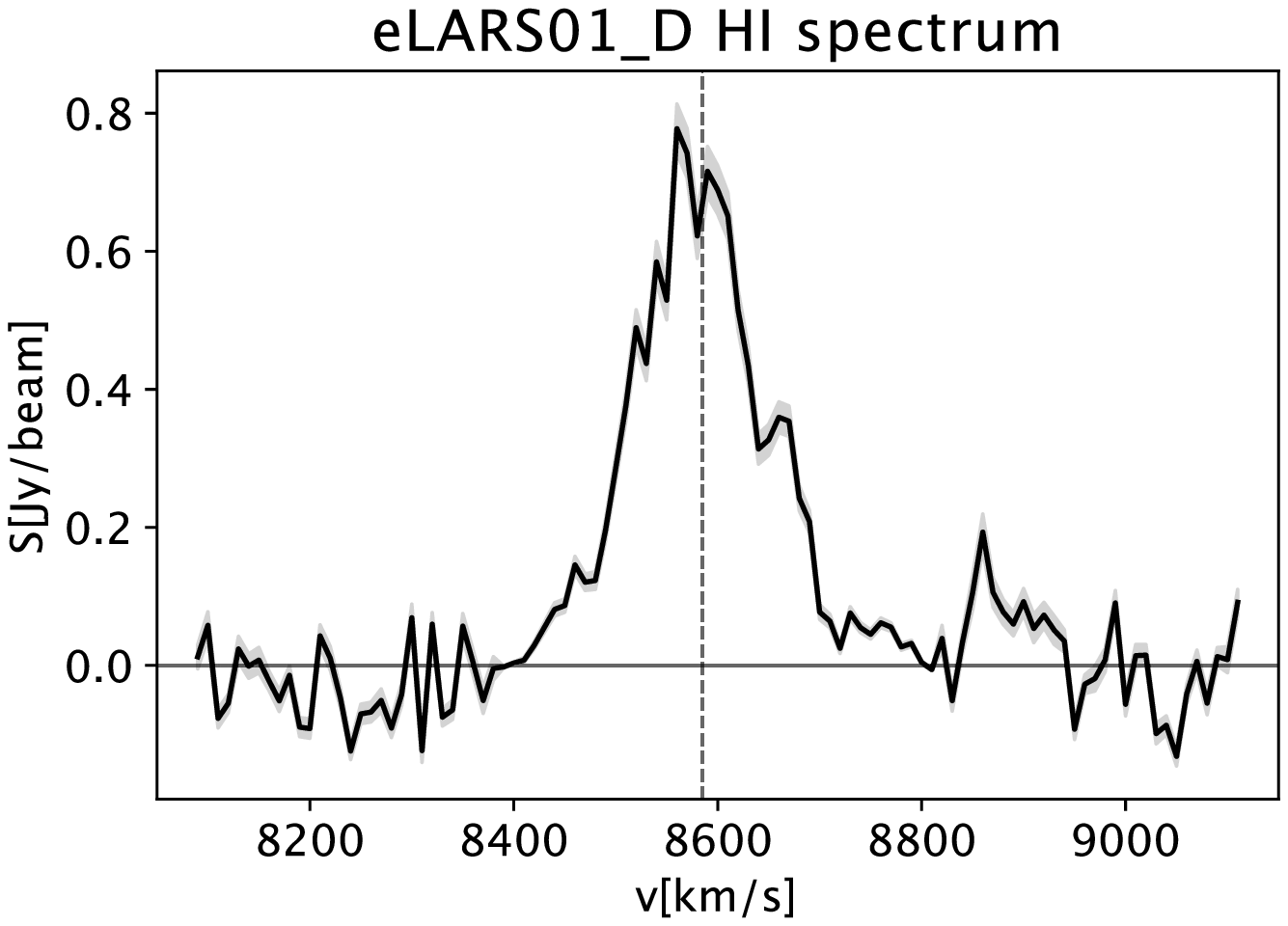}
    \caption{21cm HI emission line spectra of LARS08 and eLARS01, obtained from the masked VLA D array configuration data. The gray shaded areas show the typical error due to the flux calibration uncertainty and the cube rms. The flux in masked velocity channels was obtained from a fixed 2D aperture corresponding to the lines of sight containing emission in the cube. The vertical dotted line shows the location of the central HI velocity, the horizontal line shows the zero level. The gray area surrounding the spectrum shows the amplitude of the error due to calibrator flux uncertainty.}
    \label{fig:spec_D}
\end{figure}}

Figure \ref{fig:spec_D} shows the HI spectra of the galaxies obtained from the D configuration data. The spectrum of LARS08 shows a plateau. The associated HI flux is S$_{HI} = 1.75\pm 0.09 $ Jy.km s$^{-1}$ (with corresponding mass M$_{HI} = 1.17  \pm 0.06\times 10^{10} M_{\odot}$), in agreement within error bars with the value of S$_{HI} = 1.8 \pm 0.18$ Jy.km s$^{-1}$ found with the VLA D configuration presented in \citet[][]{Pardy2014}. As already noted in that study, there is a strong discrepancy between the flux values found using the VLA D configuration and the GBT which yields S$_{HI,GBT} = 3.4 \pm 0.3$ Jy.km s$^{-1}$. We did not find any other HI source within the area covered by the 9 arcminutes beam of the GBT, thus the difference in flux value is most likely due to the presence of low surface brightness emission that is either spatially extended and resolved out by the array or below the sensitivity limit of the VLA.
 
The spectral profile of eLARS01 is centrally peaked and we find an associated flux S$_{HI} =4.30  \pm 0.21$ Jy.km s$^{-1}$ (with corresponding HI mass M$_{HI} =1.69 \pm 0.08 \times 10^{10} M_{\odot}$). A previous HI flux measurement with the single dish Nançay radiotelescope yielded S$_{HI}=3.19$ Jy.km s$^{-1}$ \citep{Martin1991}. This telescope is much less sensitive than the VLA, with a typical rms of $\sim$1-2mJy. Error measurements on the flux were not provided in that study, however we can reasonably conclude that the lower flux value compared to the VLA is due to the large uncertainty on the flux measurement. Table \ref{tab:hi} shows the properties of the HI cubes corresponding to the different VLA configuration combinations. \\
 \begin{figure*}[!t]
  \centering
  \includegraphics[width=0.82\textwidth]{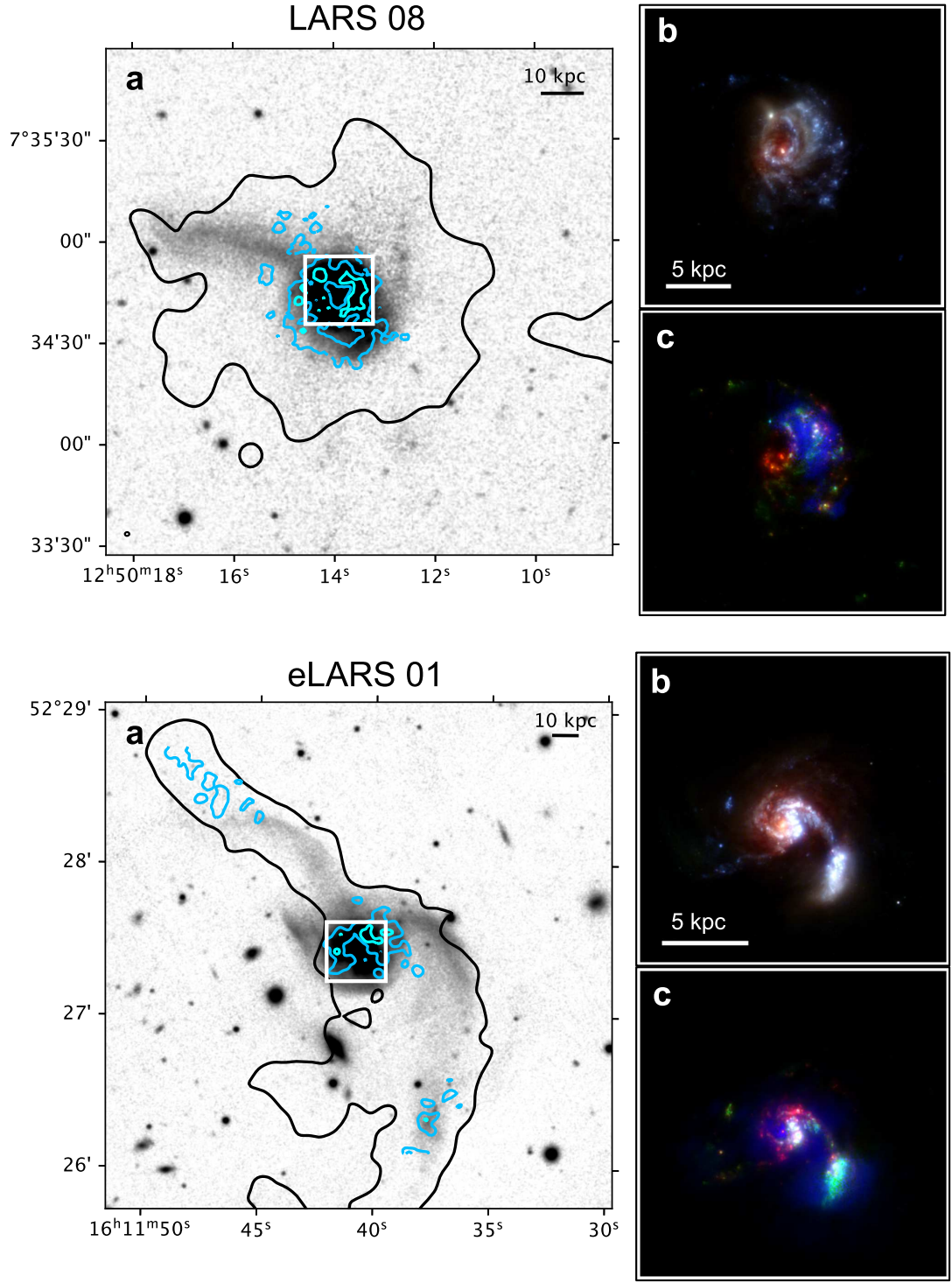}
    \caption{\textbf{a}: Pan-STARRS g,r and i stacked image (shown with asinh stretch) of LARS08 (top) and eLARS01 (bottom) with VLA D+C+B+A (cyan) and D+C (black) 21-cm HI emission contours overlaid. HI contours increase from darker to brighter. For LARS08 the contours correspond to $N_{HI}$=$5\times10^{19}$, $1\times10^{21}$ and $2.5\times10^{21}$  cm$^{-2}$. For eLARS01 the contours correspond to $N_{HI}$=$1\times10^{20}$, $1\times10^{21}$ and $3\times10^{21}$  cm$^{-2}$. The white box shows the extent of the images in \textbf{b} and \textbf{c}. \textbf{b}: HST RGB color composite image with filters F336W (blue), F438W (green), and F775W (red) for LARS08 and filters F336W (blue), F435W (green) and F814W (red) for eLARS01 \textbf{c}: HST color composite showing \lya\ (blue), \ha\ (red) and FUV (green) emission (Melinder et al., in prep.) }
    \label{fig:rgb}
\end{figure*}

Stacked Pan-STARRS and HST images with 21-cm HI contours of LARS08 and eLARS01 are presented in Figure \ref{fig:rgb}. These figures are shown for purpose of comparison between the HI and the optical emission. For both galaxies, low surface brightness features can be seen in the optical, extending beyond the HI contours. In LARS08, low surface brightness shell features in the optical indicate that some minor companion has been tidally disrupted and annexed by the main galaxy, which supports our hypothesis that the galaxy is in a group environment.\\

\subsection{\lya\ radiative transport - LARS08}
\begin{figure*}[!t]
  \centering
  \includegraphics[width=0.94\textwidth]{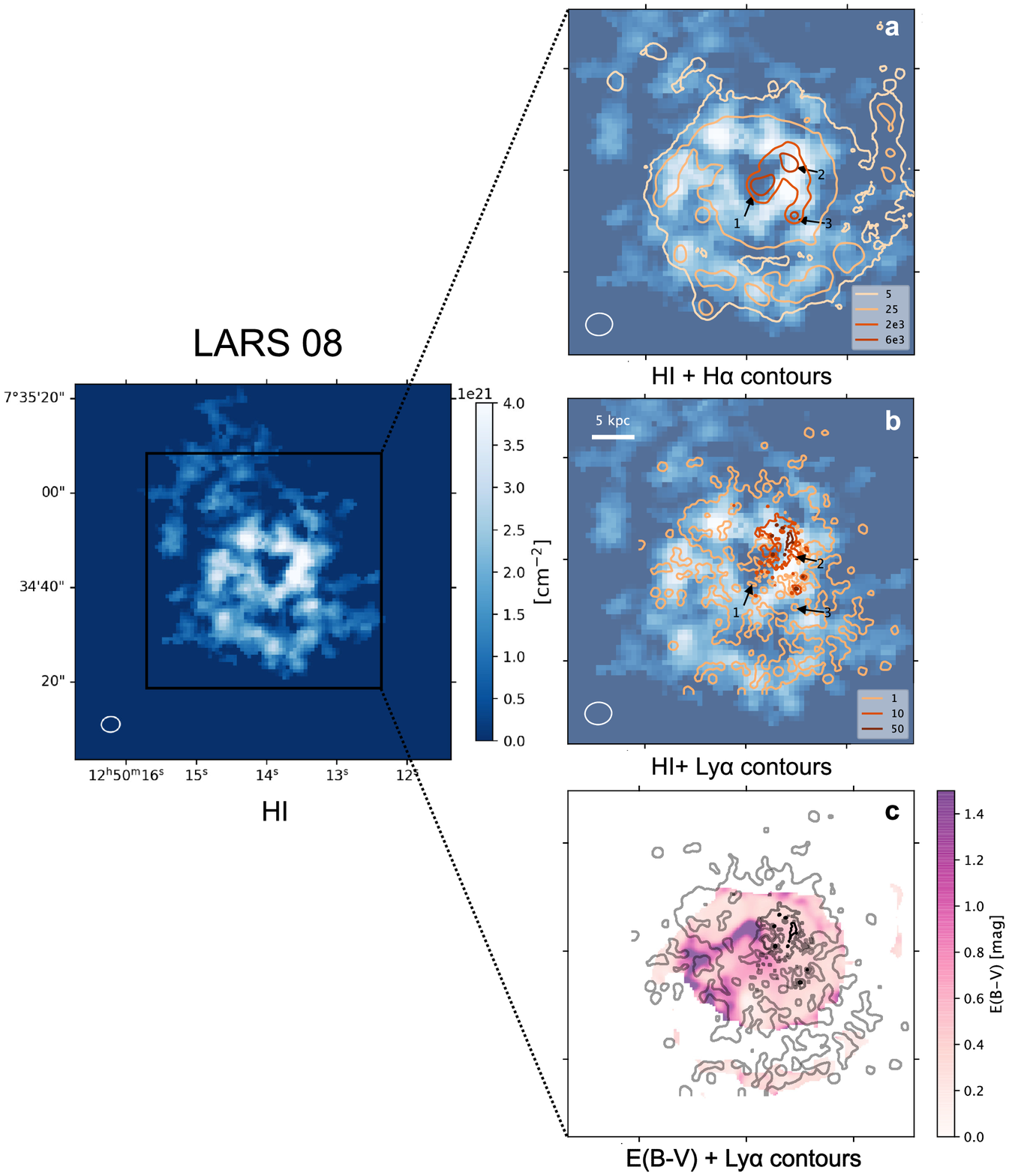}
  \caption{\textbf{Left panel:} LARS08 D+C+B+A HI moment-0 map, with box indicating the region presented on the right-hand panels overlaid in black. 
  \textbf{a}:  LARS08 D+C+B+A HI moment-0 map with MUSE \ha\ contours in units of 10$^{-18}$ erg s$^{-1}$ cm$^{-2}$ arcsec$^{-2}$ overlaid. Numbered arrows indicate the position of the strongest \ha\ regions.
  \textbf{b}: LARS08 D+C+B+A HI moment-0 map with HST \lya\ contours in units of 10$^{-16}$ erg s$^{-1}$ cm$^{-2}$ arcsec$^{-2}$ overlaid in orange. Numbered arrows indicate the position of the strongest \ha\ regions as shown on  \textbf{a}. 
  \textbf{c}: Extinction map of LARS08 with \lya\ contours overlaid in black, with levels corresponding to those in \textbf{b}.}
  \label{fig:LARS08_transfer}
  \end{figure*}

In this section and the following, we analyze the multi-wavelength observations to construct a picture of the \lya\ radiative transfer in LARS08 and eLARS01. Here, we examine ISM phases presented in Figure \ref{fig:LARS08_transfer} in turn: the \ha\ emitting regions, indicating where \lya\ is created, the HI emitting regions, showing the transport and scattering medium of \lya\ photons, and the dust, showing where \lya\ is absorbed.
  
\subsubsection{Star formation and \lya\ creation - \ha\ emission}
 Figure \ref{fig:LARS08_transfer} \textbf{a} presents contours of \ha\ emission on the HI column density map. \ha\ emission traces star-forming regions, and most importantly for \lya\ transfer, the intrinsic production locations of \lya\ radiation within the galaxy. If we were to look at a galaxy without radiative transfer effects due to HI gas or dust, the \lya\ emission morphology would thus be the same as that of the \ha. We show four \ha\ contours, corresponding to high \ha\ intensity ($>6.0\times 10^{-15}$ erg/s/cm$^2$/arcsec$^2$, red contours), intermediate \ha\ intensity ($2.5\times 10^{-17}$ erg/s/cm$^2$/arcsec$^2$ and $200\times 10^{-17}$ erg/s/cm$^2$/arcsec$^2$, orange contours) and low \ha\ intensity ($5\times 10^{-18}$ erg/s/cm$^2$/arcsec$^2$, light orange contours). Regions of high \ha\ intensity corresponding to the sites of strongest star formation are indicated by numbered arrows on this figure. The \ha\ emission  down to 5$\times 10^{-18}$ erg/s/cm$^2$/arcsec$^2$ outlines the spiral arm seen in HI. We have not represented all contours for the \ha\ emission in LARS08 to keep the \ha\ levels comparable between LARS08 and eLARS01, since the MUSE \ha\ data used for LARS08 is more sensitive than the HST data used for eLARS01. We note that the faintest MUSE \ha\ contours (down to 2.5$\times$10$^{-18}$ erg/s/cm$^2$/arcsec$^2$) reach the D+C HI contours for LARS08 and thus trace the full extent of the HI disk.

Figure \ref{fig:l08_el01_ism} in the Appendix shows, in addition to the \ha\ contours presented here, CO(2-1) contours tracing molecular gas and thus reservoirs of star-forming material. The peak of the CO emission is located in the HI hole, indicating that the hydrogen in this region is found in molecular form. The CO emission likely traces a gas reservoir for star formation, which is supported by the central high intensity \ha\ emission contours overlapping with the CO emission.
Star forming region 1 coincides with the HI hole and overlaps with the CO emission present at this location. The other two strong star forming regions that correspond to high intensity \ha\ emission are found to the West (region 2) and South-West (region 3) of the central star forming region. These high intensity \ha\ emission regions are all seen to border the highest HI column density gas (N$_{HI}>4.0\times10^{21}\,$cm$^{-2}$). The strongest star formation regions are linked by an intermediate intensity contour ($>$2.0$\times 10^{-15}$ erg/s/cm$^2$/arcsec$^2$), that overlaps with the highest HI column density gas.

\subsubsection{\lya\ transport - HI emission}
We now investigate how the \lya\ emission compares in distribution and geometry to the \ha\ and HI emissions. Figure \ref{fig:LARS08_transfer} \textbf{b} presents \lya\ emission contours overlaid on the HI column density map, with arrows indicating regions of strongest star formation presented on Fig. \ref{fig:LARS08_transfer} \textbf{a} for reference. They indicate where most of the \lya\ emission is produced, and where it would be observed in the absence of \lya\ radiative transfer effects. Three contours are overlaid on the HI map, presenting the high intensity \lya\ emission ($>5\times10^{-15}$ erg/s/cm$^2$/arcsec$^2$, dark red contour), intermediate intensity \lya\ emission ($>1\times10^{-15}$ erg/s/cm$^2$/arcsec$^2$, orange contour) and faint \lya\ emission ($>1\times10^{-16}$ erg/s/cm$^2$/arcsec$^2$, light orange contour). These contours were selected to show the  emission peaks, the intermediate \lya\ emission and the faintest contours enclosed within the HST field of view.\\

The strongest \lya\ emission ($>5\times10^{-15}$ erg/s/cm$^2$/arcsec$^2$, dark red contour) has a clumpy morphology and is found in two locations, North of star-forming region 2 and between star-forming regions 2 and 3. The strongest \lya\ emission borders, but does not coincide with, the brightest HI emission regions (N$_{HI}>4.0\times10^{21}\,$cm$^{-2}$). Instead, it is found in regions with HI column density N$_{HI}\sim3.0\times10^{21}\,$cm$^{-2}$ at the edges of HI clouds. The fact that the highest intensity \lya\ emission does not overlap with the highest HI density regions could indicate that lower spatial scattering leads to stronger \lya\ emission on the line of sight. High intensity emission would thus be seen in locations where \lya\ photons follow a path of least resistance through the neutral gas medium. However, the HI column density values N$_{HI}\sim3\times10^{21}\,$cm$^{-2}$ where the emission is seen is still very high for \lya\ radiation. We discuss this surprising result and potential projection effects in section \ref{subsec:3D_L08}. 

The intermediate intensity \lya\ emission ($>10\times10^{-16}$ erg/s/cm$^2$/arcsec$^2$, orange contour) has a more extended geometry and surrounds the strongest \lya\ emission contours. It partly overlaps with the high HI column density emission regions N$_{HI}>4\times10^{21}\,$cm$^{-2}$. This supports the hypothesis that increased spatial scattering due to higher gas column density leads to fainter \lya\ emission.
The faintest \lya\ contours ($>1\times10^{-16}$ erg/s/cm$^2$/arcsec$^2$) broadly trace the intermediate \ha\ emission ($>2.5\times 10^{-17}$ erg/s/cm$^2$/arcsec$^2$) where HI is detected in the D+C+B+A data (N$_{HI} =2.0\times10^{20}\,$cm$^{-2}$). Faint \lya\ thus reproduces the spiral structure apparent in both the \ha\ and HI data.

\subsubsection{\lya\ destruction - dust}

 Figure \ref{fig:LARS08_transfer} \textbf{c} shows the extinction E(B-V) map of LARS08, tracing the dust content of the galaxy, with \lya\ contours from Fig. \ref{fig:LARS08_transfer} \textbf{b} overlaid. Since dust absorbs \lya\ photons this effectively shows the locations where \lya\ photons are likely to be removed compared to where they are seen in emission. The E(B-V) is calculated using the \ha\ and \hb\ emission lines, assuming the extinction law from \citet{Cardelli1989} and an intrinsic line ratio of 2.86, corresponding to ionized gas with an electron density of 100 cm$^{-3}$ and temperature of 10,000 K. Although the extinction map does not cover the full extent of the HI disk due to low signal-to-noise of the weaker and more obscured \hb\ line in the outer part of the galaxy, some notable features can be seen. First, the high and intermediate intensity \lya\ emission overlaps with regions of lower extinction and dust content (E(B-V)$\lesssim$0.6), consistent with the picture where the presence of dust-rich gas would lead to \lya\ absorption. However surprisingly, the faint \lya\ emission is seen throughout the galaxy, regardless of the dust content. This suggests that regardless of extinction, sufficient spatial scattering will allow \lya\ emission to be eventually observed on the line of sight. Dust thus seems to modulate, but not prevent, extended \lya\ emission.

\subsubsection{3D analysis - the view from the COS spectra}
\label{subsec:3D_L08}
\begin{figure*}[!t]
  \centering
  \includegraphics[width=\textwidth]{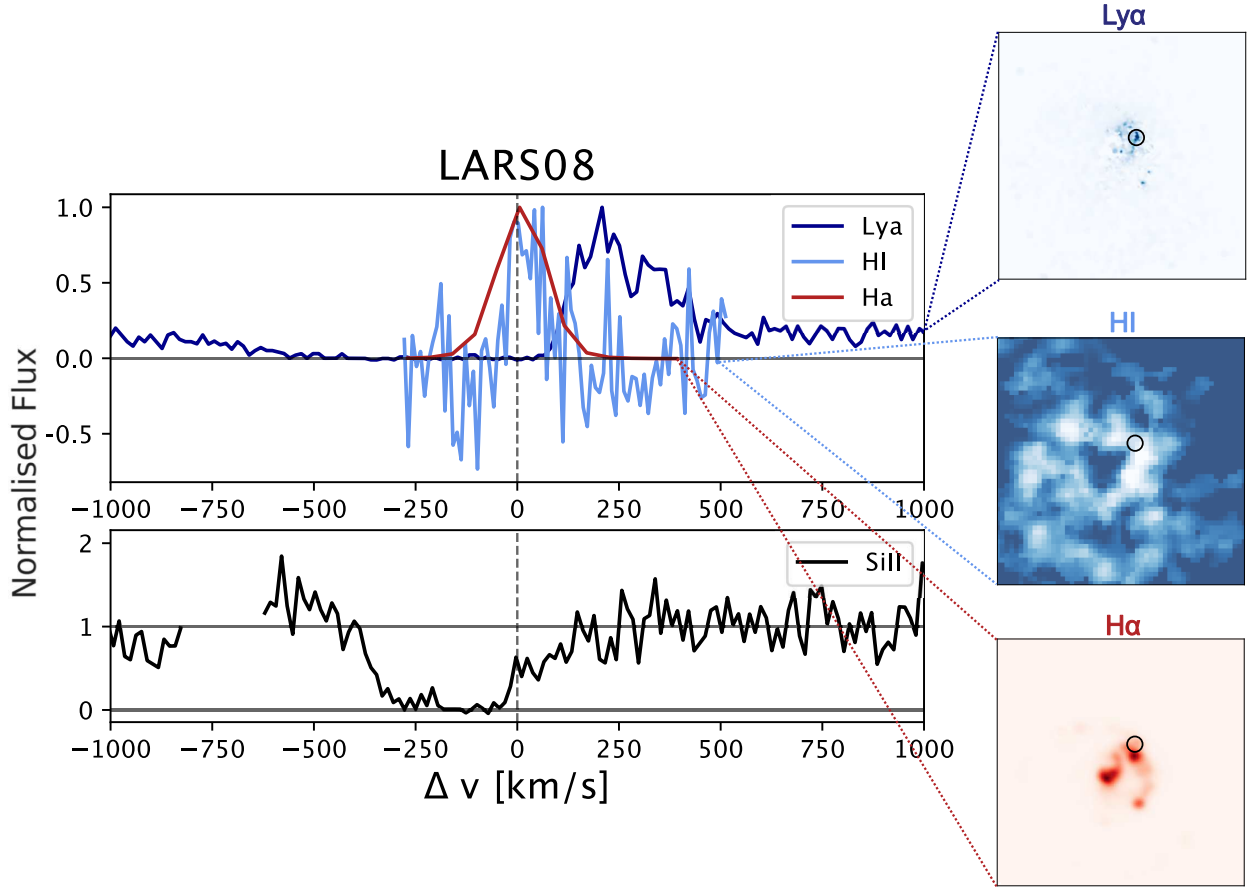}
  \caption{\textbf{Left:} Comparison of the normalised HI, \lya, \ha\ emission and \sitwo$_{\lambda 1260}$ absorption spectra as a function of the rest-frame velocity shift in LARS08. The spectra are extracted using the COS aperture, and the rest-velocity is defined by the \ha\ centroid at that position. The HI spectrum is obtained from the unmasked D+C+B+A VLA data cubes. The dashed vertical line indicates the position of the rest velocity as defined by the \ha\ centroid in the aperture, the black horizontal line indicates the zero flux level in the case of the emission spectra, and the continuum level in the case of the absorption spectrum. \textbf{Right:} Zoomed-in version of the \lya\, HI, and \ha\  intensity maps with COS aperture position represented by a black circle.}
  \label{fig:LARS08_3d}
  \end{figure*}

  Although the morphological comparison of \lya\ to the other phases of the ISM has brought insight into what impacts the spatial scattering of \lya\ photons in the galaxy, it does not solve the question of how the photons can escape from an HI medium that is extremely opaque to \lya\ (N$_{HI}\sim10^{21}\,$cm$^{-2}$ in LARS08). In this section, we make use of the HST COS spectra of the galaxies and of the 3-dimensional nature of the ionized gas IFU data and HI cubes to recover the spectral information of the various phases of the ISM presented here.\\
  
  Figure \ref{fig:LARS08_3d} shows the normalised HI and \lya\ spectra as well as the \ha\ emission and the \sitwo$_{\lambda 1260}$ absorption as a function of the radio velocity shift in the rest-frame (using the velocity of the ionized gas, determined by the optical line fitting of the MUSE spectrum at the location of the aperture) extracted using the COS aperture\footnote{We note that spectra presented in this figure are shifted compared to those presented for LARS08 in \citet[][]{Rivera2015} due to the offset position between the SDSS aperture used to recover the redshift in that previous study and that of the COS aperture. This makes the outflow seen in LARS08 less extreme compared to what had been presented previously. }. We note that the COS aperture is just 2.5~arcsec in diameter, which is of the order of the synthesized beam in the D+C+B+A data. Given that \lya\ is produced in the HII regions probed by \ha, this figure shows the relative velocity of the material that produces and scatters \lya, together with the spectral profile of \lya\ after the scattering process. The \sitwo$_{\lambda 1260}$ absorption allows one to probe the neutral medium on the same scales as \lya\ and at lower column densities (N$_{HI}$ $\geq$ 10$^{17}$cm$^{-2}$). 
  
  We observe several important features in these spectra. First, the \sitwo$_{\lambda 1260}$ absorption spectrum is blueshifted with respect to center at $\sim-140$km s$^{-1}$ (determined by moment-1), indicating an outflow of the neutral gas at that velocity. The absorption profile reaches 0, indicating high column density HI gas on the line of sight, and essentially complete covering. This is evidence that some of the neutral gas as seen in the 21cm HI map is indeed located in front of the \lya\ emission source, and is not seen to be overlapping due to projection effects.
  
 The HI emission line is overlapping with the \ha\ line, but the HI line center is slightly redshifted.
  At the location of the COS aperture, the \lya\ intensity increases as the intensity of the HI emission decreases. The whole \lya\ profile is redshifted compared to the HI and \ha\ lines and there is negligible emission at line-centre, indicating that the line must undergo a significant velocity redistribution on the red side to escape the scattering medium. The spectral shift between the HI and \lya\ lines explain how \lya\ emission can be seen to overlap spatially with high column density gas in LARS08: since \lya\ is redshifted it is effectively ``seeing" lower column density gas through the ISM. The \lya\ is seen to peak at $\sim200$ km s$^{-1}$, with a potential secondary red peak at $\sim330$ km s$^{-1}$.  While the main \lya\ peak is likely produced by scattering in the red wing of the \lya\ line, the potential presence of the secondary peak coincides with a value of $\sim$twice the expansion velocity of an ionized shell produced by the outflow, and would be expected from back-scattering on the expanding shell \citep[][]{Verhamme2006}. There is no apparent \lya\ emission on the blue side of the \lya\ line, coherent with the suppression of the blue peak by neutral outflowing gas with covering fraction of 1.\\

\subsection{ \lya\ radiative transport - eLARS01}
\begin{figure*}[!t]
  \centering
  \includegraphics[width=0.85\textwidth]{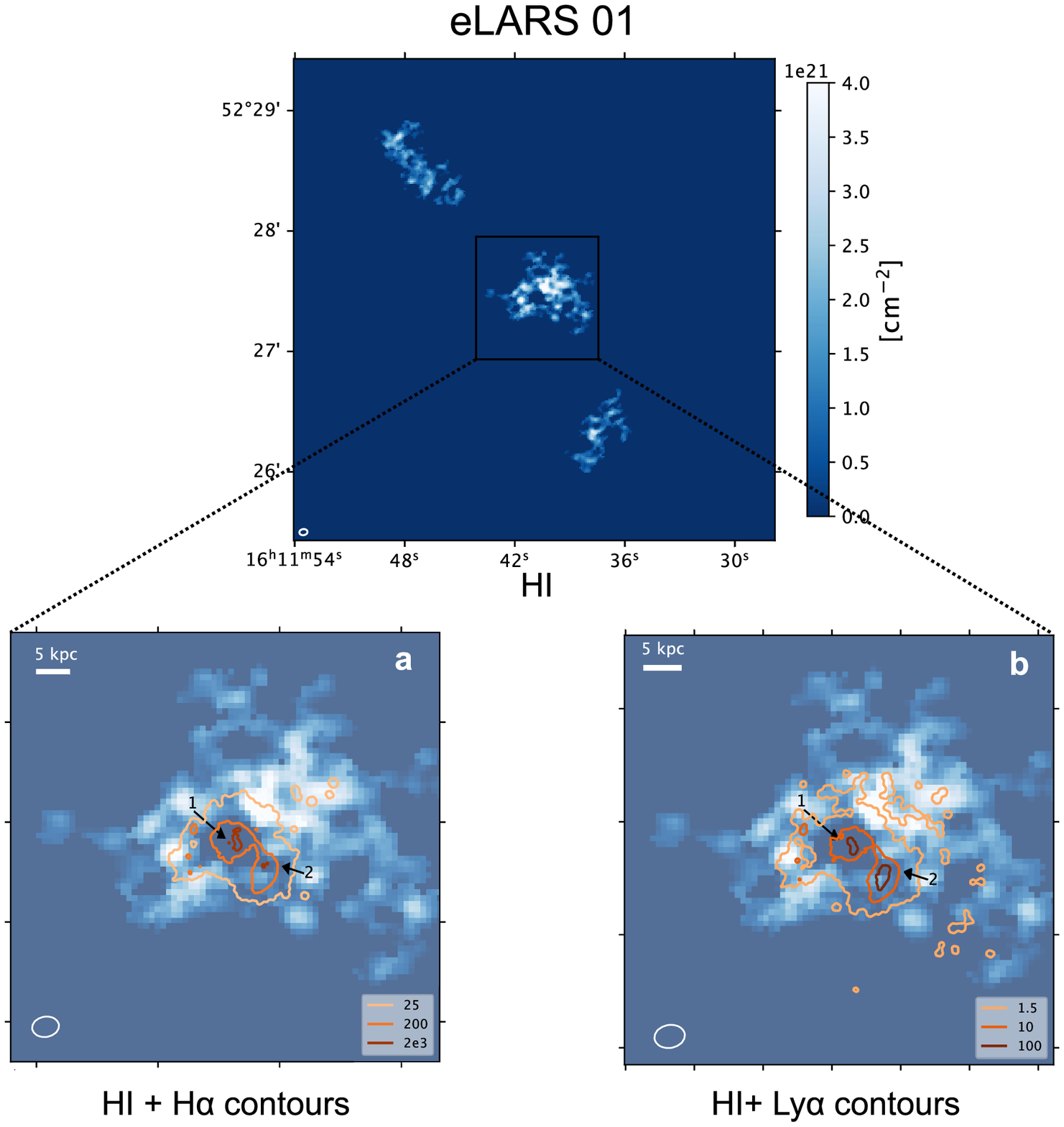}
  \caption{\textbf{Top panel:} eLARS01 D+C+B+A HI moment-0 map, with box indicating the region presented on the bottom panels overlaid in black. 
  \textbf{a}:  eLARS01 D+C+B+A HI moment-0 map with HST \ha\ contours in units of 10$^{-18}$ erg s$^{-1}$ cm$^{-2}$ arcsec$^{-2}$ overlaid in orange. Numbered arrows indicate the position of the strongest \ha\ regions.
  \textbf{b}: eLARS01 D+C+B+A HI moment-0 map with HST \lya\ contours in units of 10$^{-16}$ erg s$^{-1}$ cm$^{-2}$ arcsec$^{-2}$ overlaid in orange. Numbered arrows indicate the position of the strongest \ha\ regions as shown on \textbf{a}.} 
  \label{fig:eLARS01_transfer}
  \end{figure*}

Similarly to LARS08, we examine the ISM phases of eLARS01 presented in Figure \ref{fig:eLARS01_transfer} in turn.
\subsubsection{Star formation and \lya\ creation - \ha\ emission}
Figure \ref{fig:eLARS01_transfer} \textbf{a} shows \ha\ emission contours on the HI column density map. Numbered arrows indicate strongest \ha\ emission regions ($>$2.0$\times 10^{-15}$ erg/s/cm$^2$/arcsec$^2$). Additionally, the right panel of Figure \ref{fig:l08_el01_ism} in the appendix shows CO contours in addition to the \ha\ contours presented on Fig. \ref{fig:eLARS01_transfer} \textbf{a}. 
There are two main star forming regions in this galaxy. The high \ha\ intensity emission region to the North of the galaxy (region 1) is co-located with the 21cm absorption feature (see Figure \ref{fig:el01_abs}). The second region overlaps with an HI hole and with the peak of the CO emission. This suggests that part of the HI hole corresponds to a region where the hydrogen is found in a molecular form and fuels star formation. The optical image of eLARS01 presented in Figure \ref{fig:rgb} indicates that the \ha\ emission regions correspond to the cores of the two galaxies that are merging. Additionally, the faintest \ha\ emission contour ($\sim$2.5$\times 10^{-18}$ erg/s/cm$^2$/arcsec$^2$) extends into the gas traced by the HI maps.
Although we used a similar limiting surface brightness for our \ha\ contours, the size difference of the fields of view of MUSE used to get the \ha\ information for LARS08 and HST used for eLARS01 make it harder to quantitatively compare the \ha\ emission of eLARS01 to its extremely extended HI emission and to the \ha\ emission seen in LARS08.

\subsubsection{\lya\ transport - HI emission}
Figure \ref{fig:eLARS01_transfer} \textbf{b} shows \lya\ emission contours overlaid on the HI column density map, with regions of strongest star-formation indicated by the same arrows as presented on Fig. \ref{fig:eLARS01_transfer} \textbf{a}. In eLARS01 the strong and intermediate \lya\ emission overlaps spatially in a more direct manner with the \ha\ emission. This is possibly due to the lower column density HI gas in front of the \lya\ photon production sites (N$_{HI}<2.2\times10^{20}\,$cm$^{-2}$), leading to lower spatial scattering of the line before escape. However similarly to LARS08, the faint \lya\ contour is seen to trace the \ha\ contour at 2.5$\times 10^{-17}$ erg/s/cm$^2$/arcsec$^2$, partly following the HI emission in the D+C+B+A data. Overall, there is a very high degree of correlation between the \ha\ and \lya\ emission.\\

\subsubsection{\lya\ destruction - dust}

 We cannot obtain the dust distribution in eLARS01 in the same manner as for LARS08, due to the PMAS observational settings. Furthermore, E(B-V) maps obtained using HST imaging have lower fidelity due to the low SNR of \hb\ emission and contamination of the \ha\ by the \ntwo\ emission. The HST \ha/\hb\ map suggests that the extinction is E(B-V)$\sim$1 magnitude on average in the \lya\ emitting region, but varies between 0.2 and 2, indicating variations in the dust geometry. In this configuration, the high intensity \lya\ emission can possibly escape through the regions with lower dust content within the gas. Additionally, since the HI column density is lower, \lya\ does not scatter as much, and is thus less likely to be absorbed by dust.
 
\subsubsection{3D analysis - the view from the COS spectra}

\begin{figure*}[!t]
  \centering
  \includegraphics[width=\textwidth]{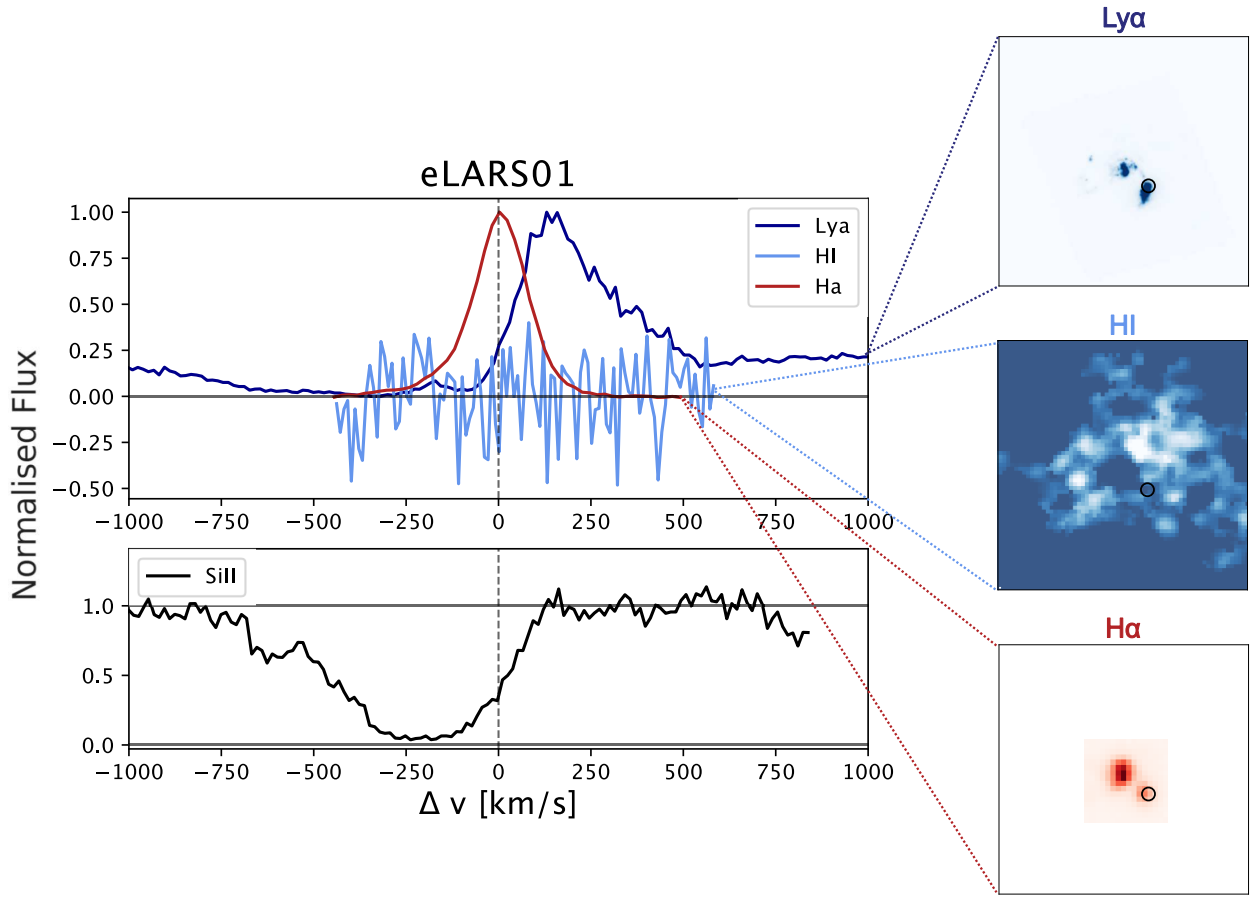}
  \caption{\textbf{Left:} Comparison of the normalised HI, \lya, \ha\ emission and \sitwo$_{\lambda 1260}$ absorption spectra as a function of the rest-frame velocity shift in eLARS01. The spectra are extracted using the COS aperture, and the rest-velocity is defined by the \ha\ centroid at that position. The HI spectrum is obtained from the unmasked D+C+B+A VLA data cubes. The dashed vertical line indicates the position of the rest velocity as defined by the \ha\ centroid in the aperture, the black horizontal line indicates the zero flux level in the case of the emission spectra, and the continuum level in the case of the absorption spectrum. \textbf{Right:} Zoomed-in version of the \lya\, HI, and \ha\  intensity maps with COS aperture position represented by a black circle.}
  \label{fig:eLARS01_3d}
  \end{figure*}
   Figure \ref{fig:eLARS01_3d} shows the normalised HI, \lya\, and PMAS \ha\ emission spectra, as well as the \sitwo$_{\lambda 1260}$ absorption spectrum as a function of the radio velocity shift in the  rest-frame extracted using the COS aperture.
  The 21cm spectrum extraction at the COS aperture in eLARS01 yields no measurable HI. The \lya\ line peaking at 160 km s$^{-1}$ is redshifted compared to the \ha\ emission, but this time some emission is seen at line center and a blue bump is seen at -170 km s$^{-1}$. The \sitwo\ absorption profile indicates an outflowing gas shell with expansion velocity at $\sim -290$ km s$^{-1}$, with the absorption extending up to -750 km s$^{-1}$. The kinematics of the \lya\ in eLARS01, in particular the smaller red peak velocity compared to LARS08 indicates that \lya\ photons need less scattering to escape the neutral gas medium, certainly due to the lower gas column density. The presence of the blue bump could be due to low covering fraction of the outflowing neutral gas in eLARS01.\\

\section{Discussion} \label{sec:discussion}
For the first time we have obtained \lya\ imaging and direct observations of the atomic gas at angular resolutions where spatial comparison with \lya\ becomes meaningful. We support these observations with optical integral field spectroscopy of the ionized gas obtained with the MUSE and PMAS instruments. Here we discuss the results of the comparison between different phases of the ISM and \lya\ emission in the observed galaxies. We recall that the \ha\ emission -- probed both spatially and kinematically by the MUSE and PMAS data -- represents the intrinsic \lya\ distribution.  After production by recombination events in the HII regions, the \lya\ photons then scatter in the HI material mapped by the VLA 21cm imaging.  This scattering process will continue until the photons either escape, in which case we might observe them in the HST imaging if they do so in the line of sight, or are absorbed by dust, which we probe for one of the galaxies using the \ha/\hb\ line ratio assuming a dust screen geometry.

\subsection{The role of mergers on \lya\ emission from galaxies}
The HI morphologies of the two galaxies presented in this study support that both are undergoing some environmental interactions. The optical image of eLARS01 clearly shows that the two galaxies of the system are currently in a major merging event that likely triggered the starburst episode. For LARS08, shell structures in the optical image indicative of such interaction were already known \citep{Micheva2018}, but the optical evidence for the infall of a galaxy is opposite to the newly detected HI tail. The connection between LARS08 and the gas cloud without optical counterpart on the Western side of the galaxy, where the gas is metal-poor (Figure \ref{fig:l08_ZN2}) could point to the accretion of metal-poor HI gas via tidal stripping, fueling the ISM of the galaxy with neutral gas. The presence of another faint HI object detected with the VLA in the vicinity of LARS08 seem to further indicate that the galaxy is located in a group environment. HI gas is very sensitive to environmental interactions, additionally more than half of the galaxies in the LARS and eLARS sample display signs of galaxy interactions or merger in the optical. This indicates that mergers might play a decisive role in fragmenting the neutral ISM and in creating the conditions that facilitate \lya\ escape. \\

\subsection{The escape of high intensity \lya\ radiation from galaxies}
We have found ISM conditions helping the escape of \lya\ photons in both galaxies.
First, we note that high star formation rates and low HI column density seem to enable the escape of high intensity \lya\ with little spatial scattering. In eLARS01, we indeed observe that the \lya\ emission morphology is very similar to the \ha\ emission, and is coincident with a location where the column density of the HI gas overlapping with \lya\ and \ha\ sites is below the detection limit ($< 2.2\times10^{20}\,$cm$^{-2}$). Our high-angular resolution observations are particularly sensitive to high-surface brightness clouds. Thus, even if not detected with the 21cm data, it is likely that some neutral gas is present in front of the \lya\ regions in a more diffuse state, or in smaller, fainter clumps. The comparison of the HI spectra also shows that the \lya\ in eLARS01 needs a smaller velocity offset compared to the \ha\ to be able to escape the galaxy. This suggests that spatial redistribution by scattering in the surrounding gas is negligible, likely due to the presence of pathways for the \lya\ emission to escape directly. Furthermore, the COS \lya\ spectrum shows a blue bump peaking around v$\sim$-170 km s$^{-1}$, which should be suppressed by the outflow of neutral gas.  In order for the \lya\ emission to escape and have such morphology, a relatively low dust content mixed with the HII gas in the regions where we see \lya\ is also needed, although we could not evaluate this precisely with the available data. However, the \ha/\hb\ line ratio obtained from HST imaging suggests a very heterogeneous dust medium, with escape through dust-free channels possible thanks to the lower HI column density which limits spatial scattering along the line of sight.\\
High star formation rates, high HI column density, and low dust contents on the line of sight also enable the escape of high intensity \lya\ emission under certain gas kinematics. In LARS08, the highest intensity \lya\ emission is clumpy and is found around regions of intense star formation that border the highest column density HI clouds in regions with low dust content (Figure \ref{fig:LARS08_transfer}). The fact that the high intensity \lya\ emission does not overlap with the highest intensity gas clouds despite the intense star formation taking place on the line of sight suggests that \lya\ photons follow paths of least resistance when scattering through the gas by avoiding the densest clouds. However, it is surprising that \lya\ can escape such dense medium at all. Absorption in the \lya\ spectrum (Figure \ref{fig:LARS08_3d}) is visible along the line-of-sight out to velocities of almost -1000 km s$^{-1}$, which implies significant absorption of the UV continuum along the line-of-sight, with column
densities in excess of 10$^{20}$ cm$^{-2}$, which is optically thick
to \lya ($\tau\sim10^5$). Furthermore, on the line of sight where the COS spectrum is obtained, the \sitwo\ absorption line reaches zero, implying a covering fraction of 1. 
Figure \ref{fig:LARS08_3d} shows the \lya\ is strongly redshifted with respect to rest velocity defined by nebular lines in the MUSE spectrum. The \lya\ emission peaks at a velocity of $\sim$200 km s$^{-1}$, where co-spatial \ha\ decreases significantly, but where the HI density is significantly lower. Furthermore the \lya\ emission as seen in the COS spectrum extends to significantly redder velocities, well beyond where detectable amounts of HI are found. Therefore, \lya\ photons seem to scatter in velocity space in order to ``bypass" the high column density HI gas on the line of sight and escape on the red side. We conjecture that the mechanism responsible for the velocity-space scatter of the \lya\ line at the COS aperture position is a combination of scattering in the red wing of the line and back-scattering of the \lya\ photons on the HI material of an expanding medium. Indeed examining the neutral gas in the COS aperture through \sitwo\ line absorption, one finds gas moving out to high velocities ($-$140  km s$^{-1}$ on average and $-$365 km s$^{-1}$ at maximum), guaranteeing the presence of a strong outflow. This is not seen in the HI data, however \sitwo\ probes column densities $\geq10^{17}$ cm $^{-2}$ and can be found both in HII and HI gas, and hence is not tracing exactly the same medium as the 21-cm line is. The potential presence of a secondary peak would be expected from back-scattering of the \lya\ photons on the expanding shell.
The velocity redistribution of \lya\ photons is probably facilitated by the low dust obscuration on this line of sight, meaning that \lya\ can undergo many scattering events with limited absorption, leading to high intensity emission. On the other hand, no strong or intermediate intensity \lya\ emission is seen overlapping with star forming region 1 despite the high star formation intensity, probably due to the high dust obscuration (E(B-V)$>0.6$), which seems to significantly dampen the intensity of the emission at this location. Thus dust content seems to impact the escape of high intensity \lya\ emission out of galaxies. \\

\subsection{Faint \lya\ emission and HI structure}
We now discuss our results regarding the faint \lya\ emission. We find that in both galaxies, the faint \lya\ emission (1-1.5  10$^{-16}$ erg/s/cm$^2$/arcsec$^2$)  broadly follows the \ha\ emission at $5-25\times 10^{-18}$ erg/s/cm$^2$/arcsec$^2$ where HI is detected with the D+C+B+A data (N$_{HI} \geq2\times10^{20}$cm$^{-2}$). In LARS08, the faint \lya\ traces the spiral arm, regardless of dust content on the line of sight. Thus while dust seems to impede high intensity \lya\ emission, it does not prevent the formation of low intensity \lya\ halo. \citet[][]{Hayes2013} found that the size of \lya\ halos in LARS galaxies was strongly correlated with quantities that also scale with dust content. This suggests that dust modulates \lya\ emission by either dampening the \lya\ output or allowing only the escape of \lya\ emission from unobscured star forming regions, resulting in a faint \lya\ halo. We note that the \lya\ emission contours we can access are limited by the HST field of view. Thus our data cannot answer the question of how fainter \lya\ is distributed relative to the neutral gas and whether it traces the extent of the HI in the galaxies we study here.

\section{Conclusions and Outlook} \label{sec:conclusions}

We have presented the first comparative study of \lya\ and 21-cm HI emission at high angular resolution in galaxies, supported by molecular and ionized gas observations. This study of two low-z starburst galaxies shows the observational link between \lya\ creation, scattering and destruction and the local structure of the interstellar medium. The main results of our study are summarized hereafter:
\begin{itemize}
    \item[-] Both galaxies are undergoing interactions which has possibly played a role in triggering the starburst episodes they are experiencing. In eLARS01 a major merger is ongoing, while LARS08 is interacting with a companion object with no optical counterpart. The companion is either a low-surface brightness dwarf galaxy or a tidal remnant which has been imaged here for the first time. Additional deep optical imaging reveals shell features indicative of a prior merger.  
    \item[-] We find that high-intensity \lya\ emission ($>5\times 10^{-15}$ erg/s/cm$^2$/arcsec$^2$) is found overlapping with regions of intense star formation traced by the \ha\ emission. We find two ISM states facilitating the emergence of high-intensity \lya\ emission. In eLARS01, with low column density HI on the line of sight, high intensity \lya\ emission can escape with little velocity redistribution and the \lya\ morphology is very similar to that of the \ha\ emission. In LARS08, we observe that high intensity \lya\ emission can escape in the presence of high column density HI with low dust extinction, but the line needs significant velocity redistribution on the red side.
    \item[-]  Fainter \lya\ emission ($1-1.5\times 10^{-16}$ erg/s/cm$^2$/arcsec$^2$) broadly traces regions of intermediate \ha\ intensity where we observe HI in the D+C+B+A data. For the galaxy where we can measure dust obscuration, it is present regardless of dust content on the line of sight, implying that increased scattering will eventually lead to the formation of a faint \lya\ halo. In this context dust obscuration seems to modulate, but does not impede the formation of large-scale \lya\ emission. Lower intensity \lya\ emission might trace HI structures on larger scales but the limited field of view of HST compared to the 21cm data does not allow us to assess the full scale of the \lya\ emission.
\end{itemize}

The results we have obtained highlight the importance of the intermediate-scale ($\sim$kpc) ISM structure to \lya\ escape, in particular high intensity \lya\ emission. Although the analysis has remained very exploratory, we have been able to explain \lya\ escape in both galaxies using simple scenarios that suggest that scaling relations between HI, \ha, dust and \lya\ could be found.\\

The two galaxies presented here are interesting case studies, but they represent a very small sub-sample of the LARS+eLARS galaxies, so a quantitative analysis will require a larger sample size. Obtaining further VLA A-configuration observations for the 15 galaxies already having intermediate angular scale observations (D+C+B) would be a way to extend the results presented here to a larger range of galaxies and ISM environments. However, the use of the highest angular resolution data presents challenges in terms of exposure time and signal-to-noise ratio, that make this approach impractical for a large sample of galaxies. Additionally, we identified that the combination of D, C, and B data allows to identify several of the HI substructures and represents a compromise between angular resolution and sensitivity (see Figures \ref{fig:moment_lowres_L08} and \ref{fig:moment_lowres_eL01}). Thus in future work, we will exploit the VLA B-configuration data already obtained for 15 LARS galaxies to explore the link identified here between \lya\ escape and resolved ISM properties. Some future work will also include detailed analysis of the HI emission properties and kinematics of the full LARS + eLARS galaxies at lower angular resolution (VLA D+C data), to quantify the impact of global HI properties on \lya\ emission. These will be important steps to quantify the impact of neutral gas on \lya\ radiative transfer at a variety of physical scales.

\acknowledgments
A.L.R. would like to thank the referee for their comments and members of the Stockholm University Astronomy Department galaxy group for their feedback throughout the project.\\
M.H. is fellow of the Knut and Alice Wallenberg Foundation.\\
E.C.H thanks the staff at Calar Alto Observatory for help with the visitor mode observations.\\
G.O. was supported by the  Swedish  Research  Council (Vetenskapsr\r{a}det).\\
J.M.C. and N.V. acknowledge support from Macalester College.
DK is funded by the  CNES/CNRS agreement N. 180027.
The National Radio Astronomy Observatory is a facility of the National Science Foundation operated under cooperative agreement by Associated Universities, Inc.\\
Based on observations made with the NASA/ESA Hubble Space Telescope, obtained from the data archive at the Space Telescope Science Institute. STScI is operated by the Association of Universities for Research in Astronomy, Inc. under NASA contract NAS 5-26555.\\
Based on observations collected at the European Southern Observatory under ESO programme 0101.B-0703(A).
Data here reported were acquired at Centro Astronómico Hispano Alemán (CAHA) at Calar Alto operated jointly by Instituto de Astrofísica de Andalucía (CSIC) and Max Planck Institut für Astronomie (MPG). Centro Astronómico Hispano en Andalucía is now operated by Instituto de Astrofísica de Andalucía and Junta de Andalucía.
\vspace{5mm}
\facilities{VLA, HST(ACS,COS), VLT:Yepun(MUSE), PMAS, IRAM:NOEMA, SMA, ALMA }
\vspace{5mm}
\software{Astropy \citep[][]{astropy2013, astropy2018}, casa \citep[][]{casa2007} }
\vspace{5mm}

\bibliography{bibli}{}
\bibliographystyle{aasjournal}

\newpage
\appendix
\section{Lower-angular resolution HI maps of LARS08 and eLARS01}
\begin{figure*}[!h]
    \centering
    \includegraphics[width=0.79\textwidth]{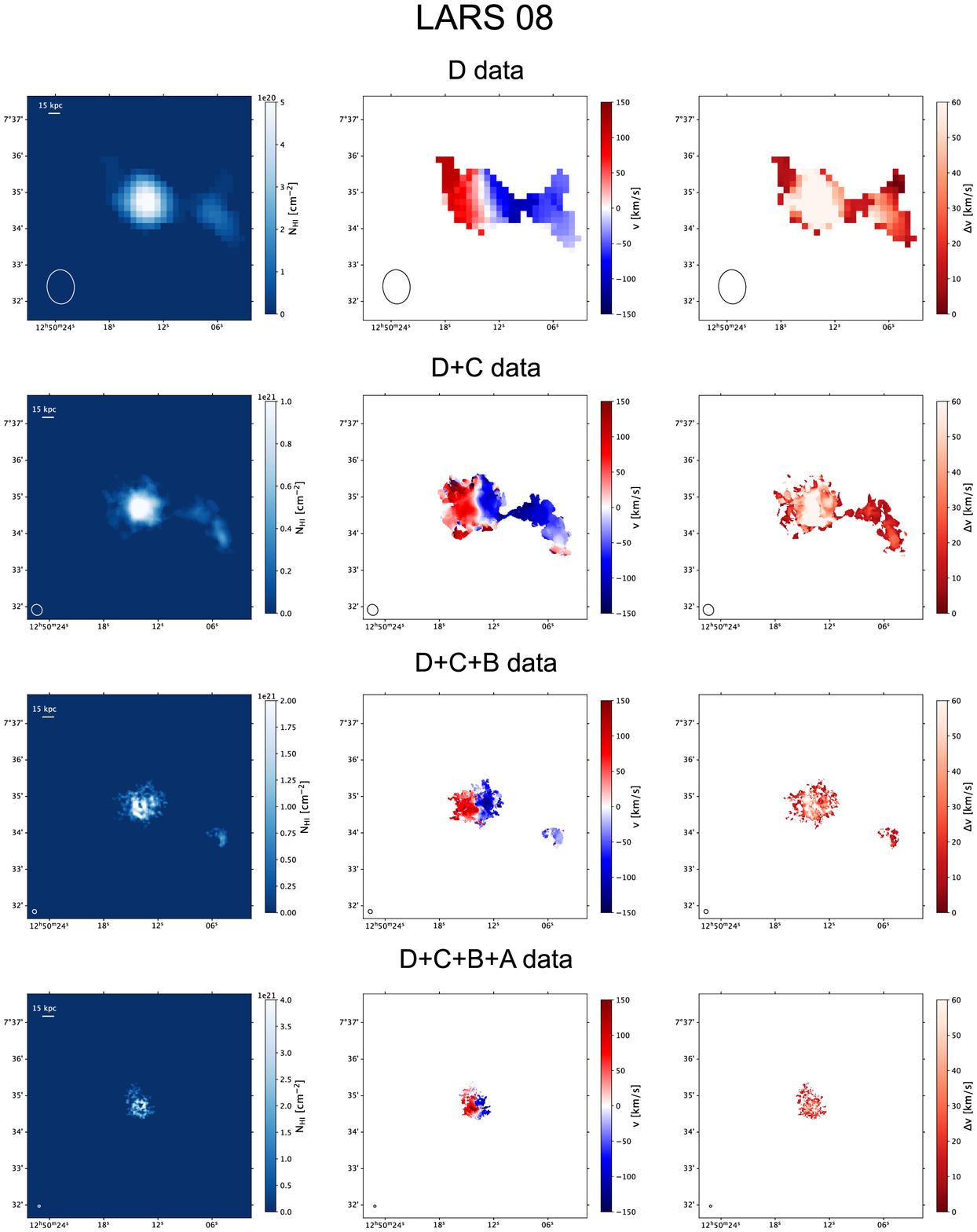}
    \caption{Moment maps for the 21cm HI emission line of LARS08, obtained with the VLA D, D+C, D+C+B and D+C+B+A configuration data. From left to right: column density map, rest-frame intensity-weighted velocity (moment-1) map, velocity dispersion map.}
    \label{fig:moment_lowres_L08}
\end{figure*}
\begin{figure*}[!htb]
    \centering
    \includegraphics[width=0.79\textwidth]{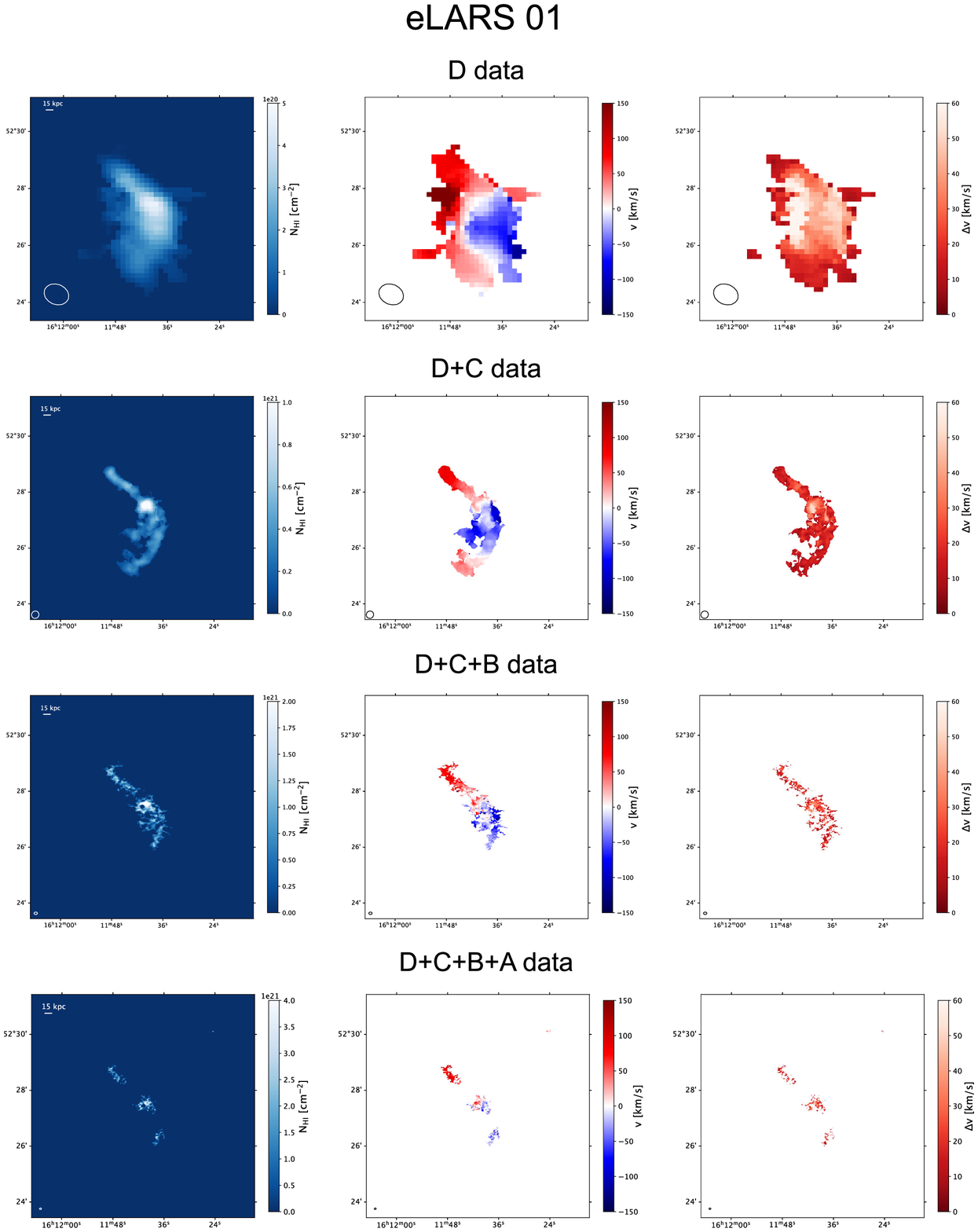}
    \caption{Moment maps for the 21cm HI emission line of eLARS01, obtained with the VLA D, D+C, D+C+B and D+C+B+A configuration data. From left to right: column density map, rest-frame intensity-weighted velocity (moment-1) map, velocity dispersion map.}
    \label{fig:moment_lowres_eL01}
\end{figure*}

\newpage
\section{HI channel maps}
\begin{figure*}[!h]
    \centering
    \includegraphics[width=0.79\textwidth]{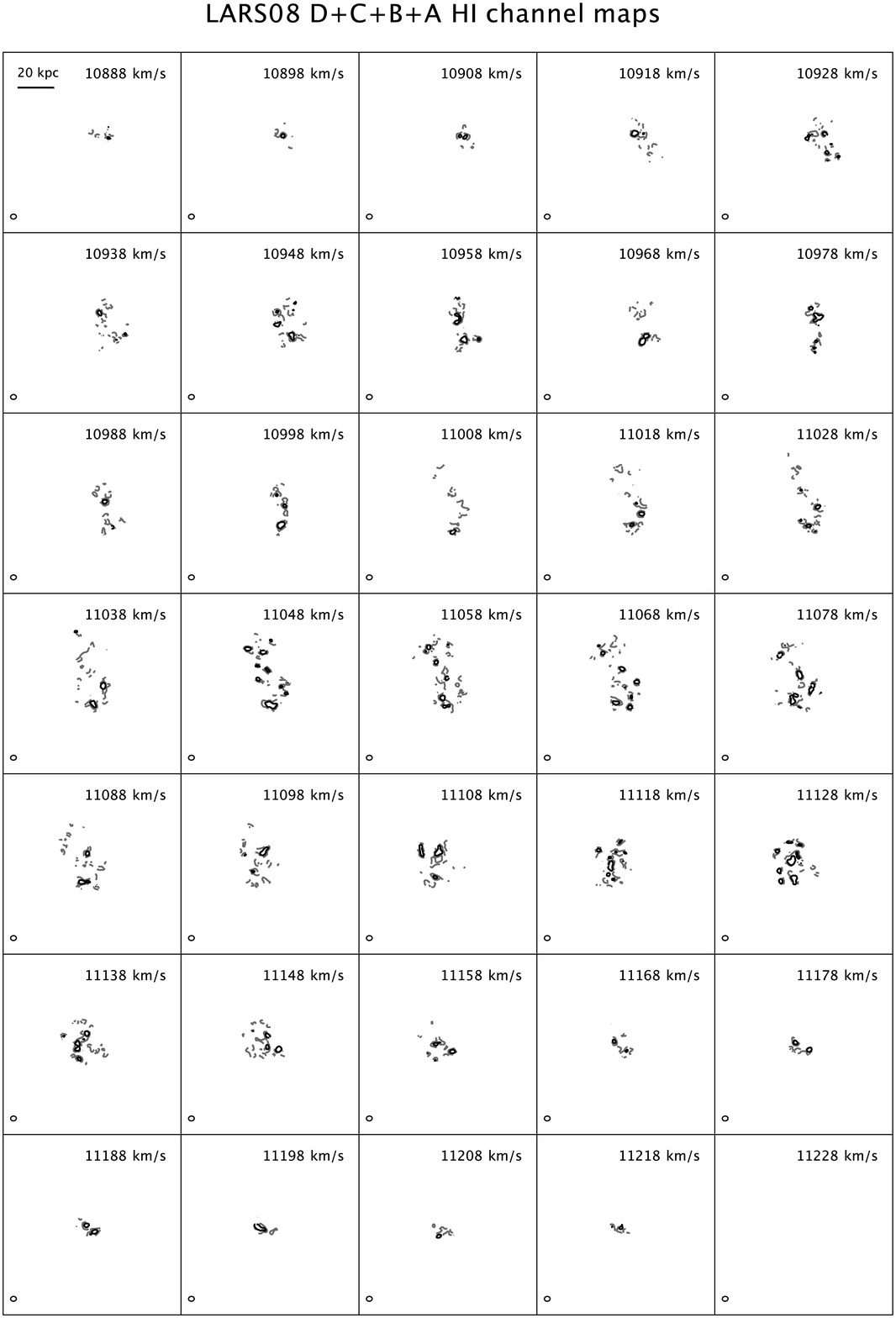}
    \caption{Full HI channel map for the D+C+B+A data of LARS08 with contours of 0.24 mJy/beam in grey and 0.4 mJy/beam in black corresponding to the 1.5$\sigma$ and 3$\sigma$ noise levels.}
    \label{fig:l08_hi_cm_full}
\end{figure*}
\begin{figure*}[!tbhp]
    \centering
    \includegraphics[width=0.75\textwidth]{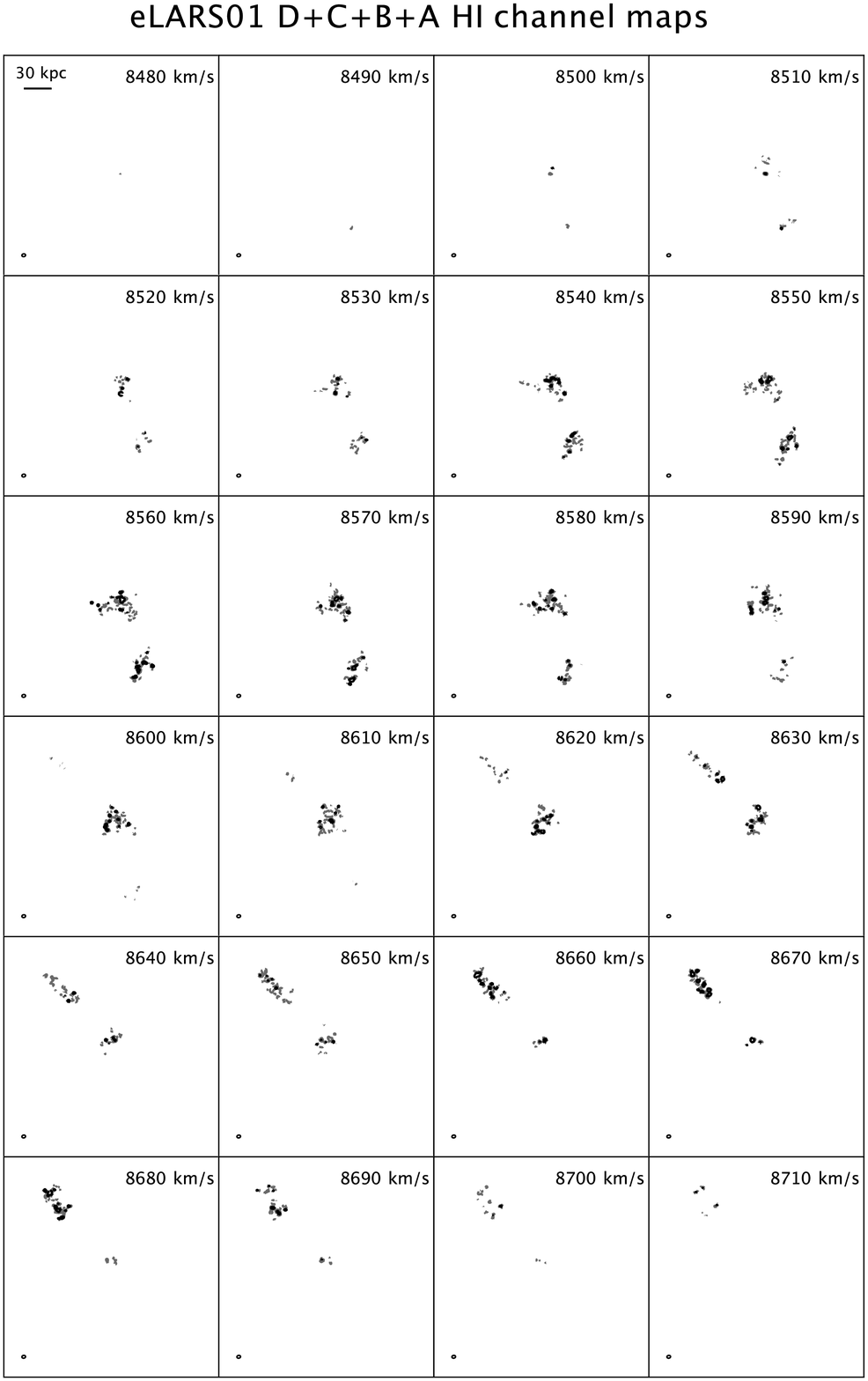}
    \caption{Full HI channel map for the D+C+B+A data of eLARS01, with contours of 0.26 mJy/beam in grey and 0.51 mJy/beam in black corresponding to the 1.5$\sigma$ and 3$\sigma$ noise levels.}
    \label{fig:el01_hi_cm_full}
\end{figure*}

\newpage

\section{21cm absorption in eLARS01}
\begin{figure*}[!h]
    \centering
    \includegraphics[width=0.8\textwidth]{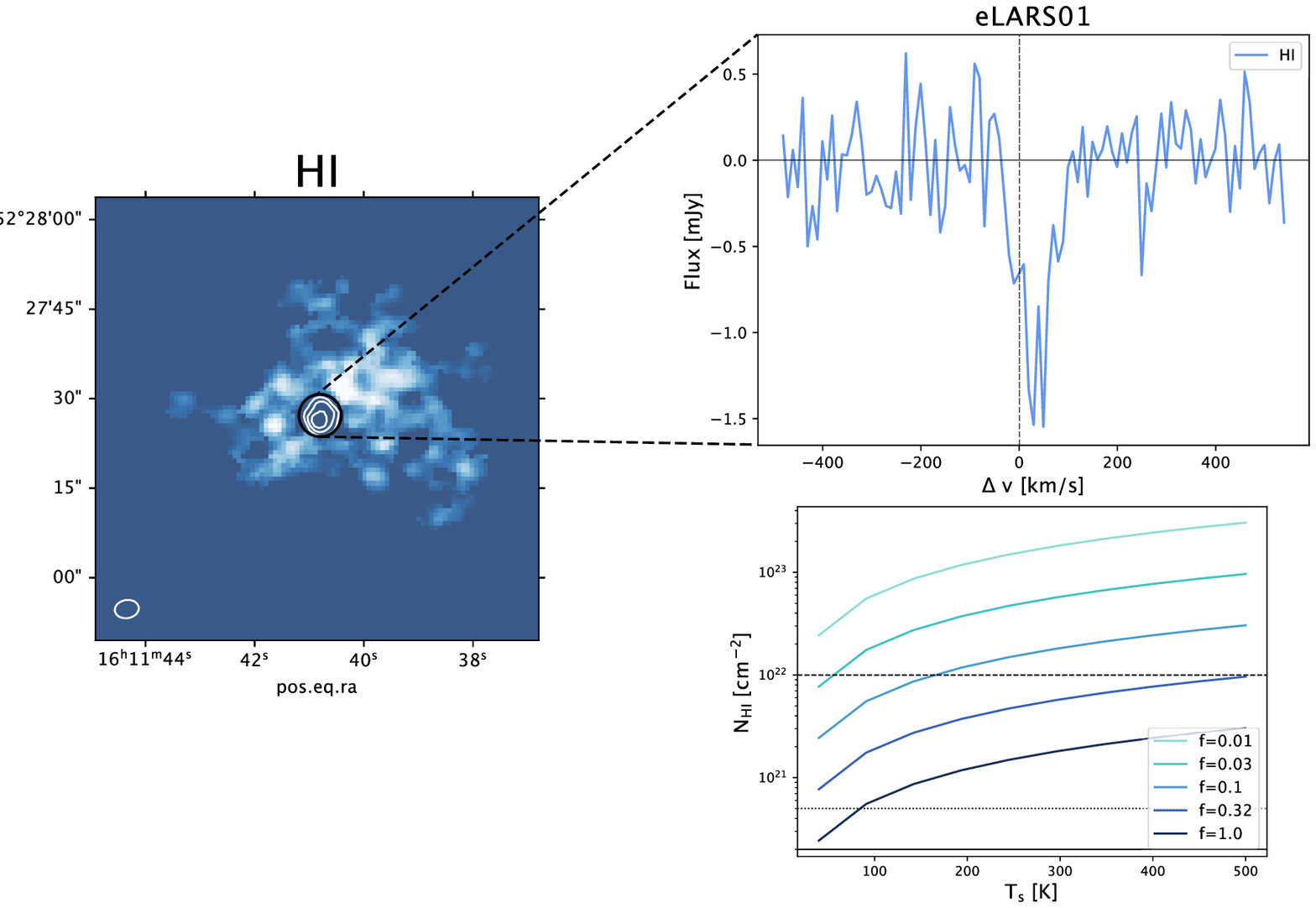}
    \caption{\textbf{Left panel:} 21cm HI moment-0 map with absorption contours overlaid in white in units of [0.5, 1.0, 1.5] $\times10^{18}\,Ts/f$ cm$^{-2}$ . The aperture used to extract the spectrum is shown by a black circle. \textbf{Top right panel:}  21cm absorption profile extracted from the aperture shown on the left panel. \textbf{Bottom right panel:} Total 21cm absorbing column density as a function of spin temperature T$_s$ and covering fraction $f$ for temperatures characteristic of the cold neutral medium. The solid line corresponds to the observed self-shielding limit of HI \citep{Kanekar2011}, the dotted line corresponds to the threshold for the formation of molecular hydrogen \citep{Savage1977} and the dashed line to the threshold where most of the neutral gas is molecular \citep{Schaye2001}.}
    \label{fig:el01_abs}
\end{figure*}
Figure \ref{fig:el01_abs} shows the absorption profile extracted from an aperture the average size of the beam in eLARS01. The velocity offset has been calculated using the \ha\ centroid at the aperture location. The HI gas seen in absorption is redshifted compared to the nebular gas, indicating an inflow of neutral gas towards the star-forming region. Although a \lya\ spectrum is not available at this location, the presence of the neutral gas inflow suggests that we would observe a \lya\ profile with a strong blue peak at the location of 21cm absorption. It is interesting to find evidence of a neutral gas inflow overlapping with the Northern star forming region in eLARS01, just 5 kpc away from the second star forming region where the COS spectrum was obtained, and where the \sitwo\ absorption profile indicates a neutral gas outflow. This suggests that the feedback processes occurring in the two star-forming regions are different. 

The absorbing column density in cm$^{-2}$ can be calculated using $N_{HI} = 1.822 \times 10^{18}\ T_s/f \, \frac{F_{HI,abs}}{F_{HI,cont}}\, \Delta v$, where $T_s$ is the spin temperature in K, $f$ is the covering fraction, $F_{HI,abs}$ is the absorbing flux in mJy and  $F_{HI,cont}$ the continuum flux in mJy. We find that the continuum flux at the 21cm frequency is  $F_{HI,cont}$ = 28.9$\pm$0.9mJy. The significant uncertainty on the spin temperature T$_{s}$ and the covering fraction $f$ limits our ability to calculate the column density of the absorbing gas. On the bottom right panel of Figure \ref{fig:el01_abs} we show the column density as a function of the spin temperature and covering fraction for  T$_{s}$ values characteristic of the cold neutral medium (40-500K, \citet{Kanekar2011}). We find that the absorbing gas is above the effective self-shielding limit for HI $N_{HI} >2\times10^{20}$cm$^{-2}$  \citep{Kanekar2011}, consistent with the gas being observed in front of a star-bursting location emitting UV emission. Without self-shielding the HI at this location would likely be ionized. Since the peak of the absorption feature is found just next to the location where molecular gas is observed (see Figure \ref{fig:l08_el01_ism}), we conclude that the column density is likely above the threshold $N_{HI}=5\times10^{20}$cm$^{-2}$ where HI starts to become molecular \citep{Savage1977}, but below  $N_{HI} <1\times10^{22}$cm$^{-2}$, the threshold for which hydrogen is found in a predominantly molecular form \citep[][]{Schaye2001}. These limits poorly constrain the range of values for the spin temperature and the covering fraction, however the similarity between the \ha\ and \lya\ morphologies at this location indicate that \lya\ photons escape with relatively little scatter through the neutral gas of the galaxy. This could be explained by the neutral gas having a low covering fraction, which is not excluded by the limits, with covering fractions as small as 3\% being possible for a spin temperature T$_{s}<55\,$K.

\section{CO in LARS008 and eLARS01}
\begin{figure*}[h]
    \centering
    \includegraphics[width=0.45\textwidth]{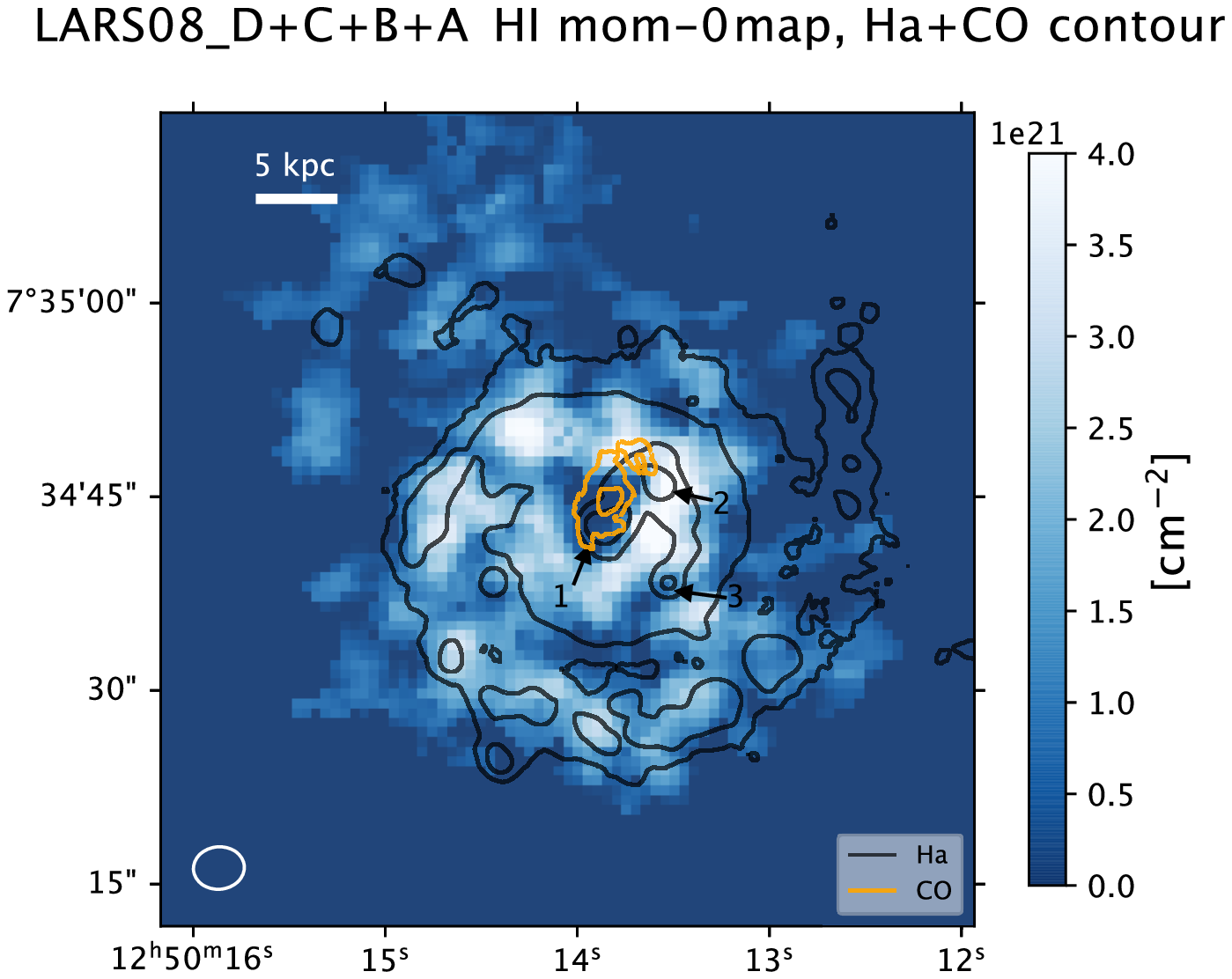}
    \includegraphics[width=0.45\textwidth]{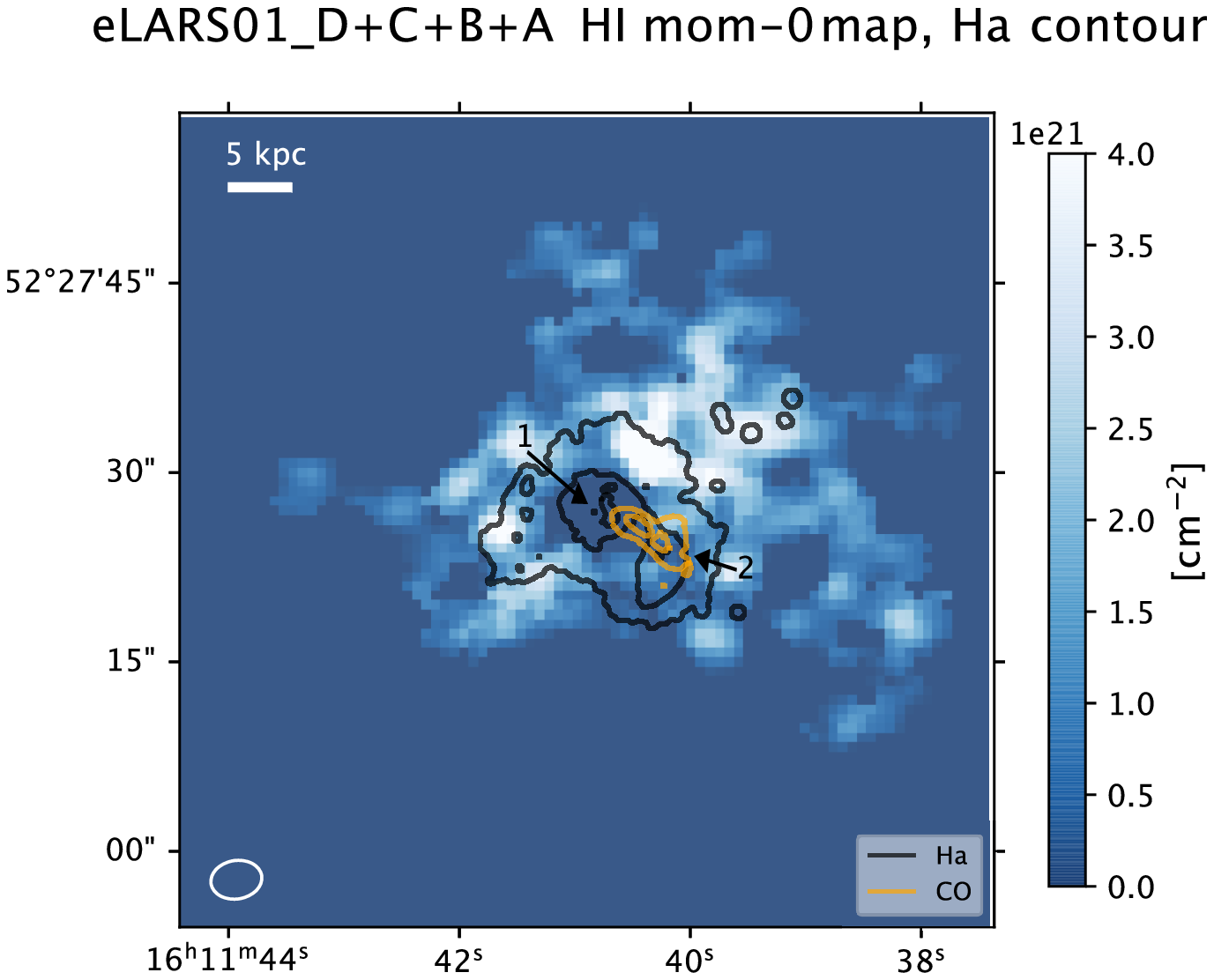}
    \caption{Zoomed-in portion of the HI column density map, with MUSE \ha\ contours overlaid in black and CO(2-1) contours in yellow. For LARS08 (left), the \ha\ contour levels increase from the outer part of the galaxy, with the same levels as in Fig. \ref{fig:LARS08_transfer} \textbf{a} displayed. For eLARS01 (right), the \ha\ contour levels increase from the outer part of the galaxy, with the same levels as in Fig. \ref{fig:eLARS01_transfer} \textbf{a} displayed. CO(2-1) contours representing the extent and peak of the emission are shown in arbitrary units. In each panel, arrows labelled with numbers indicate the position of the regions with strongest \ha\ emission. The size of the VLA D+C+B+A synthesized beam is indicated on the bottom left of the  mom-0 map by a white ellipse.}
    \label{fig:l08_el01_ism}
\end{figure*}

In Figure \ref{fig:l08_el01_ism}, we show \ha\ and CO(2-1) contours on the HI D+C+B+A moment-0 maps of LARS08 and eLARS01 (The CO data will be presented in Puschnig et al, in prep.). Since CO(2-1) traces molecular gas, the CO contours show star-forming regions. CO contours overlap with regions of strong star formation and low HI column density in both galaxies, suggesting the drop in HI column density is linked to the presence of Hydrogen in a molecular, rather than atomic form. 
\section{Metallicity map of LARS08}
\begin{figure*}[h]
    \centering
    \includegraphics[width=0.48\textwidth]{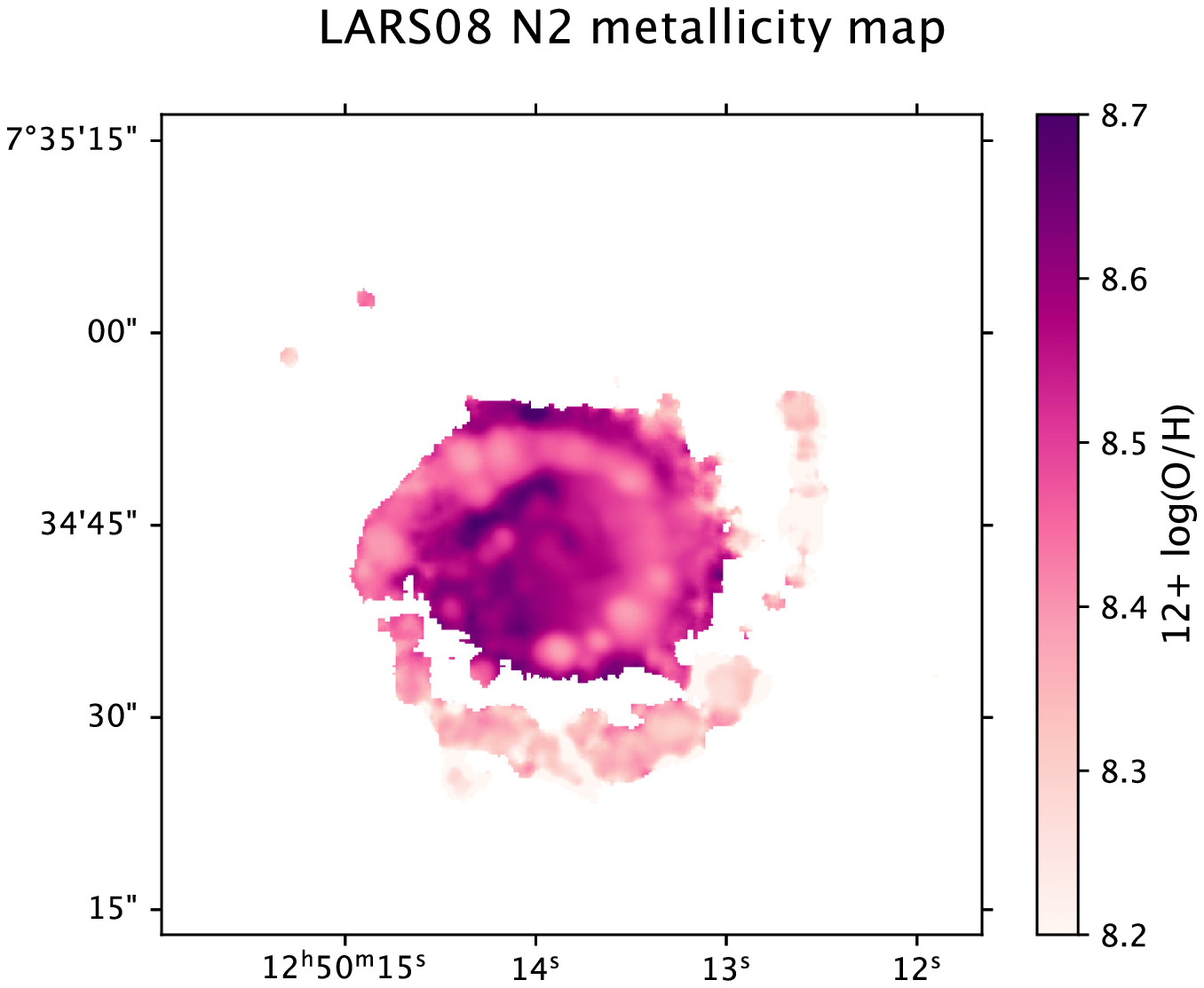}
    \caption{N2 metallicity map of LARS08.}
    \label{fig:l08_ZN2}
\end{figure*}
Figure \ref{fig:l08_ZN2} presents the N2 metallicity map of LARS08, derived from the [N$_{II}$]$_{\lambda 6584}$/\ha\ ratio in the MUSE data cube using the \citet{Marino2013} calibration. The metallicity exhibits a radial gradient, and is lower along the main spiral arm, especially in the outer part to the South and West of the galaxy. The strongest metallicities are found in the inner part of the galaxy, to the East in particular.

\end{document}